\documentclass[aps,prb,longbibliography,notitlepage,nofootinbib,onecolumn,superscriptaddress,12pt,tightenlines]{revtex4-1}

\pdfoutput=1

\usepackage{graphicx,dcolumn,bm,amsmath,amssymb,euscript}

\usepackage[utf8]{inputenc}

\usepackage[colorlinks,citecolor=red,urlcolor=blue]{hyperref}

\usepackage{color, colortbl}
\definecolor{LightCyan}{rgb}{0.88,1,1}

\graphicspath{{Figures/vsubmit/}}

\usepackage{miller}   

\begin{document}

\title{Persistence of Island Arrangements During Layer-by-Layer Growth Revealed Using Coherent X-rays}

\author{Guangxu Ju}
\author{Dongwei Xu}
\altaffiliation[current address: ]{School of Energy and Power Engineering, Huazhong University of Science and Technology, Wuhan 430074, China}
\author{Matthew J. Highland}
\affiliation{Materials Science Division, Argonne National Laboratory, Argonne, IL 60439 USA}
\author{Carol Thompson}
\affiliation{Department of Physics, Northern Illinois University, DeKalb IL 60115 USA}
\author{Hua Zhou}
\affiliation{X-ray Science Division, Argonne National Laboratory, Argonne, IL 60439 USA}
\author{Jeffrey A. Eastman}
\author{Paul H. Fuoss}
\altaffiliation[current address: ]{SLAC National Accelerator Laboratory, Menlo Park, CA 94025 USA}
\author{Peter Zapol}
\affiliation{Materials Science Division, Argonne National Laboratory, Argonne, IL 60439 USA}
\author{Hyunjung Kim}
\affiliation{Department of Physics, Sogang University, Seoul, 04107 Korea}
\author{G. Brian Stephenson}
\email[correspondence to: ]{stephenson@anl.gov}
\affiliation{Materials Science Division, Argonne National Laboratory, Argonne, IL 60439 USA}

\date{April 22, 2018}

\begin{abstract}
Understanding surface dynamics during epitaxial film growth is key to growing high quality materials with controllable properties.
X-ray photon correlation spectroscopy (XPCS) using coherent x-rays opens new opportunities for \textit{in situ} observation of atomic-scale fluctuation dynamics during crystal growth.  
Here, we present the first XPCS measurements of 2D island dynamics during homoepitaxial growth in the layer-by-layer mode. 
Analysis of the results using two-time correlations reveals a new phenomenon -- a memory effect in island nucleation sites on successive crystal layers.
Simulations indicate that this persistence in the island arrangements arises from communication between islands on different layers via adatoms.
With the worldwide advent of new coherent x-ray sources, the XPCS methods pioneered here will be widely applicable to atomic-scale processes on surfaces.
\end{abstract}

\maketitle

Nanoscale structures generated by growth processes during phase transitions are ubiquitous in synthesis of advanced materials, and are often central to their outstanding properties.
These include systems as diverse as
precipitates and twin textures in high strength and shape-memory alloys \cite{2013_Song_Nature502_85,2016_Ogawa_Science353_368,2015_Gornostyref_PCCP17_27249},
spatio-temporal patterns in surface reconstructions and nanoparticles driven by catalytic chemistry \cite{2006_Mikhailov_PhysRep425_79,2008_Jiang_PRL101_086102},
and domain patterns in ferroic thin films that generate high dielectric, conductivity, piezoelectric, and magnetic responses \cite{2016_McLeod_NatPhys13_80,2017_Huang_NatRevMat2_17004}.
To understand the competing fundamental processes that control nucleation and growth of these structures, we need to observe more than just the average nucleation rate or density, but also the preferred sites and arrangements.
X-ray scattering methods are well suited for directly observing the formation of such structures \textit{in situ} during materials synthesis, although traditional methods with incoherent x-rays typically provide only spatially averaged quantities.
The advent of coherent x-ray methods has opened a new window into phase transition mechanisms, by revealing the dynamics of the exact nanostructure arrangement and its fluctuations \cite{shpyrko2014x}.

Standard x-ray photon correlation spectroscopy (XPCS) analysis characterizes the time correlation of fluctuations in equilibrium or steady-state systems.
In many cases, the correlations simply decay exponentially, and the wavenumber $(Q)$ dependence of the correlation time provides direct information about the mechanism of the fluctuation dynamics \cite{pierce2009surface, kim2003surface, ruta2014revealing, roseker2018NatComm}.
In other cases, heterodyne mixing between scattering from different regions produces oscillatory correlations, allowing the relative velocities of different structural features to be determined \cite{ulbrandt2016direct}.

When the average structure is not constant in time, e.g. during domain nucleation and growth \cite{PiercePRL2003, SanbornPRL2011, chesnel2011oscillating, chesnel2013field, chesnel2016shaping, pierce2005disorder}, coarsening \cite{MalikPRL1998, Livet2001kinetic, fluerasu2005x, kim2016synchrotron}, or structural relaxation \cite{wang2015free, ruta2017hard}, a two-time XPCS analysis can be applied \cite{bikondoa2017use}.
For domain coarsening processes, such analysis has shown that the domain arrangement can be remarkably independent of time, with the random pattern established by the initial nucleation process simply amplified or diminished at different $Q$ by coarsening \cite{MalikPRL1998, brown1997speckle, brown1999evolution, fluerasu2005x}.
Nucleation of magnetic domains during field cycling has been shown to exhibit a memory effect \cite{PiercePRL2003, chesnel2011oscillating, chesnel2013field, chesnel2016shaping,pierce2005disorder}, with strong correlations between the domain arrangements occurring on each cycle of the field.

Here we explore the spatio-temporal correlations of two-dimensional (2D) nanostructures produced in a different class of phase transition - crystal growth from the vapor.
During crystal growth in the layer-by-layer growth mode, deposited atoms diffuse on the surface to nucleate islands of single monolayer (ML) height, which grow and coalesce to form a complete layer \cite{neave1983dynamics,tsao2012materials}.
This process of 2D island nucleation and coalescence repeats in a cyclic fashion to form each layer of the crystal.
Current models predict that the arrangement of islands reflects a competition between nucleation of new islands versus attachment to existing islands, kinetically controlled by the relative rates of deposition and adatom diffusion, and affected by the Ehrlich-Schwoebel (ES) barrier for diffusion over step edges \cite{kaufmann2016critical}.
To investigate these fundamental phenomena, we have chosen to study metal-organic vapor phase epitaxy (MOVPE) of GaN on its 
{$(1 \, 0 \, \overline{1} \, 0)$}
m-plane surface, a growth orientation of interest for advanced solid-state lighting and high power electronics \cite{denbaars2013development}.
Previous \textit{in situ} x-ray scattering studies with incoherent x-rays \cite{perret2017island} have characterized the layer-by-layer growth mode in this system, and mapped the average island spacing and shape.
Using XPCS, we have discovered that the island arrangements on each layer can be highly correlated, persisting over several monolayers of growth.
To understand the cause of this persistence, we have performed kinetic Monte Carlo (KMC) simulations of the growth process.
These show a similar persistence in the island arrangements between layers, indicating that the effect is not due to nucleation at fixed defects as in magnetic domain systems \cite{pierce2005disorder}, but is instead due to communication from layer to layer via the adatom density distribution.

\begin{figure*}
\includegraphics[trim=50 50 50 50,clip,width=\linewidth]{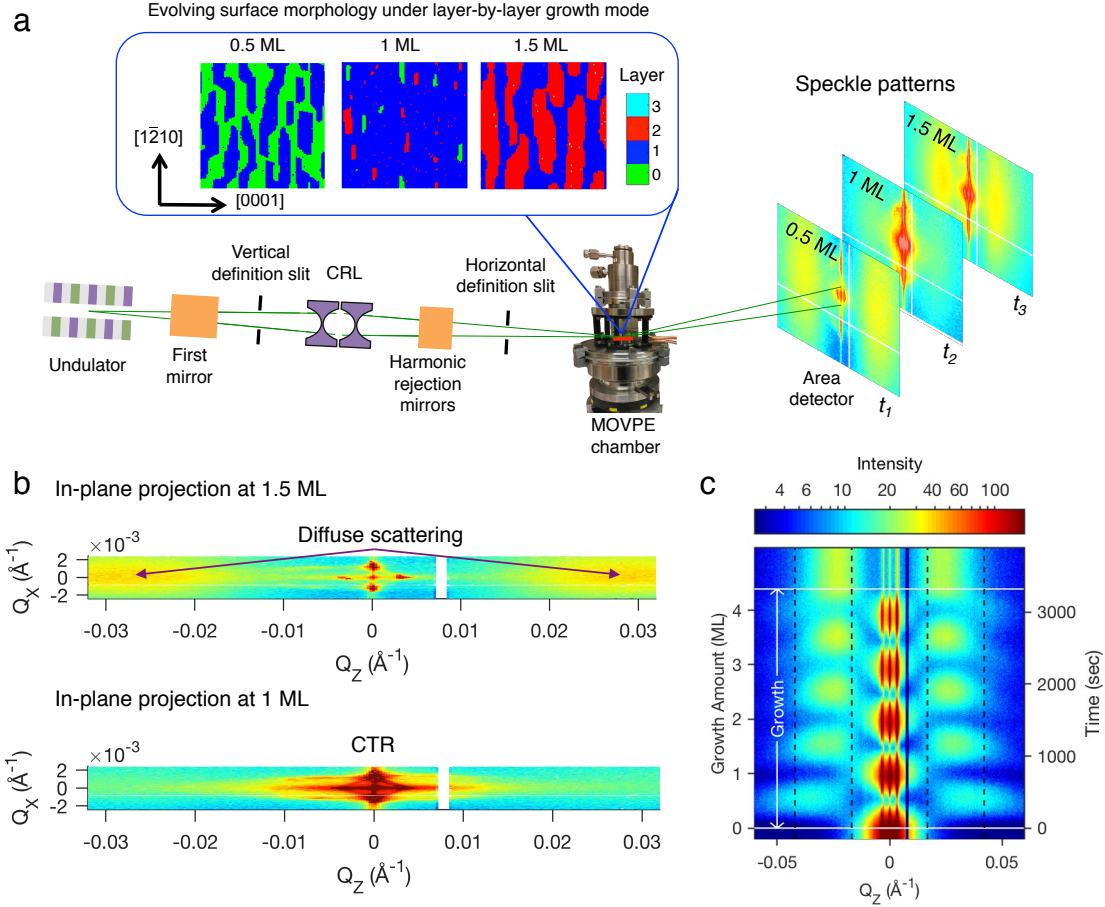}
\caption{\textbf{Experimental schematic and diffuse scattering from islands.} 
\textbf{a}, The evolving arrangement of atomic-height islands during metal-organic vapor phase epitaxy (MOVPE) on the GaN 
{$(1 \, 0 \, \overline{1} \, 0)$}
m-plane surface is recorded as a function of time in the speckle pattern produced by a partially coherent x-ray beam prepared by a special optical setup (Supplementary Section 1).
\textbf{b}, Scattering patterns recorded near the $Q_Y = 1.09 $~\AA$^{-1}$ position along the specular direction are projected onto the $Q_X$-$Q_Z$ surface plane in reciprocal space, so that features can be directly related to the surface morphology. 
(White lines near $Q_Z = 0.008$ and $Q_X = -0.001$~\AA$^{-1}$ are artifacts from the missing pixels between the detector quadrants.)
In the typical images shown, diffuse scattering at $Q_Z \approx 0.025$~\AA$^{-1}$ arises from 2D islands with spacings of $\sim25$ nm, while crystal truncation rods (CTRs) at low $Q_Z$ arise from $\sim10$ $\mu$m facets having $\sim100$ nm atomic terraces (Supplementary Fig.~1 and Section 2).
Speckle in the diffuse scattering allows analysis of the spatio-temporal correlations in the island arrangement. 
\textbf{c}, Intensity integrated between $Q_X = \pm 3\times10^{-4} $~\AA$^{-1}$ as a function of time during growth of 4.5 monolayers (1~ML = 0.276~nm) in the layer-by-layer mode at 796~K.
Times of growth start and end are indicated.
During growth, the CTR intensity oscillates with time, with minima and maxima at half- and full-integer-ML amounts of growth, respectively, owing to destructive interference between scattering by each successive layer at this $Q_Y$.
The diffuse scattering intensity oscillates out of phase with the CTRs, with maxima at half-integer ML when island density is largest, and minima at full-integer ML when islands have coalesced to form a new layer.
       }
\label{fig:Overview}
\end{figure*}

\section*{XPCS Measurements}

Because of the low scattering cross section of 2D islands, developing XPCS methods to study their dynamics presents significant challenges.
To date, XPCS studies of atomic-height surface features have been performed only on strongly scattering systems such as Au with relatively large ($\sim$1 $\mu$m) in-plane structures \cite{pierce2009surface,pierce2012dynamics}.
The higher x-ray energy ($E \approx 25$ keV) required for \textit{in situ} studies, to penetrate the environment and to avoid interaction of the x-ray beam with the growth process, creates an additional challenge since both the coherent x-ray flux available from the source and the solid angle required to resolve speckle at the detector scale as $E^{-2}$.
To increase the transversely coherent flux available from the source, we removed the monochromator and accepted the full pink beam bandwidth $(\Delta E / E = 1.3 \times 10^{-2})$ of the third harmonic of the undulator spectrum.
While this relatively large bandwidth decreases the $Q$ resolution in the radial direction, it can be adequate for near-specular XPCS measurements from surface features, since the speckle are extended in the surface normal direction (Supplementary Fig.~4).
A special x-ray optical setup was developed to optimize the coherence lengths and remove unwanted harmonics from the undulator spectrum (Fig.~\ref{fig:Overview}(a) and Supplementary Section 1).
We also developed correlation analysis techniques optimized for low signals.  

Another major challenge in performing an \textit{in situ} XPCS study of growth is to maintain position and angular stability of the sample at high temperature so that the incident beam illuminates the same area during the growth process. 
We developed a special chamber and instrument for coherent x-ray studies during \textit{in situ} growth by MOVPE \cite{ju2017instrument}, 
and verified that it had the required stability (Supplementary Fig.~5 and Section 3).

During homoepitaxial growth of GaN on the 
{$(1 \, 0 \, \overline{1} \, 0)$}
m-plane surface, the islands that form are elongated perpendicular to the 
{$[0 \, 0 \, 0 \, 1]$}
direction because of step edge energy anisotropy \cite{perret2017island,xu2017kinetic}.
This concentrates the diffuse scattering into the $Q_Z$ direction (Fig.~\ref{fig:Overview}(b)).
We monitored the speckle pattern in the diffuse scattering from islands near the specular rod 
{$(H \, 0 \, \overline{H} \, 0)$}
at $H = 0.48$, close to but just off the anti-Bragg position because of the residual harmonic contamination at $H = 0.5$. 
The intensity of the diffuse scattering from the islands oscillates out of phase with the crystal truncation rod (CTR) intensity (Fig.~\ref{fig:Overview}(c)), as expected for layer-by-layer growth \cite{perret2017island}. 
A slow growth rate of $\sim1.3\times 10^{-3}$ ML/s was used to increase signal collection time,
and temperatures studied were chosen to span across the layer-by-layer growth regime \cite{perret2017island}.

Because the system is not at steady-state during layer-by-layer growth and the average intensity is oscillating with time, we analyzed the speckle pattern sequence using a two-time correlation function 
for wavevector $\textbf{Q}$ and times $t_1$ and $t_2$,
\begin{equation}
C(t_{1},t_{2})
    = \left  \langle  
        \frac{\Delta I(\textbf{Q},t_{1})\Delta I(\textbf{Q},t_{2})}
        {\overline{I}(\textbf{Q},t_{1})\overline{I}(\textbf{Q},t_{2})}
            \right \rangle_\textbf{Q},
\label{eq:ttc}
\end{equation}
where $\Delta I(\textbf{Q},t_i) \equiv I(\textbf{Q},t_i) - \overline{I}(\textbf{Q},t_i)$ is the deviation of the intensity in the speckle pattern from the mean intensity $\overline{I}$ that would be measured under incoherent conditions where the speckles are not resolved. 
For analysis of our experimental results, where the mean intensity $\overline{I}$ varies with time, we obtain $\overline{I}(\textbf{Q},t)$ by smoothing $I(\textbf{Q},t)$ over a range of neighboring detector pixels, and we obtain the ensemble average $\langle ~ \rangle$ by averaging over a range of $\textbf{Q}$ having similar time correlations (Supplementary Table 1 and Section 3).
This form of the two-time correlation function gives values which are analogous to the contrast of a distribution, since the diagonal elements with $t_1 = t_2$ are equal to the observed variance divided by the square of the mean.
The measured speckle contrast values are in good agreement with those expected from the properties of the x-ray illumination and the detector (Supplementary Figs.~2, 3).
Values of $C$ greater than, equal to, or less than zero correspond to island arrangements which are correlated, uncorrelated, and anti-correlated, respectively.

\begin{figure*}
\includegraphics[trim=0 300 0 0,clip,width=\linewidth]{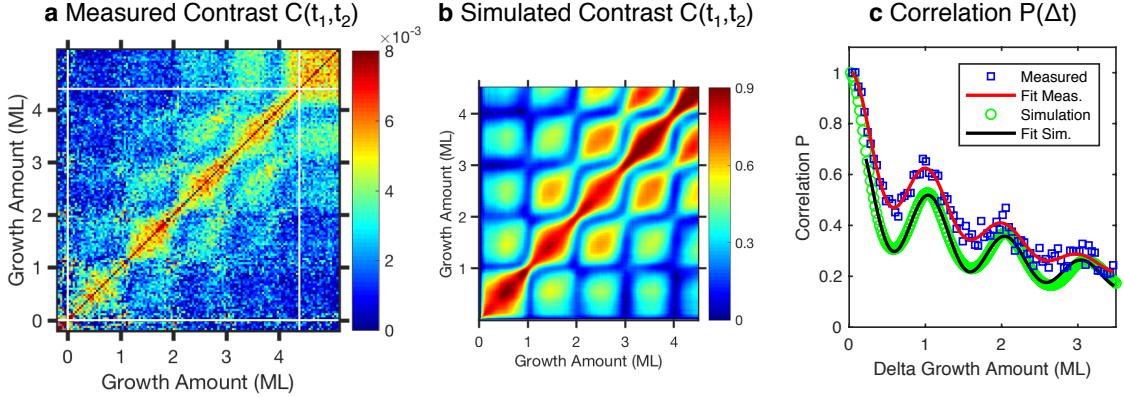}
\caption{\textbf{Oscillating correlations.} \textbf{a}, Two-time correlations in the measured speckle from 2D islands before, during, and after growth at 796 K. The time axes have been converted to growth amount using the observed growth rate (Supplementary Table 2). White lines indicate the start and end of growth. The ``checkerboard'' pattern indicates that the arrangement of islands formed in successive layers is correlated with that in previous layers. \textbf{b}, Two-time correlations in the simulated speckle from 2D islands during growth, which also show correlations between island arrangements in successive layers. \textbf{c}, Points: measured and simulated correlations averaged over equal time differences, showing oscillatory behavior with a period of 1 ML. Curves: fits to equation~(\ref{eq:fit}).}
\label{fig:Diffuse_2time}
\end{figure*}

To characterize the 2D island arrangements, we extract two-time correlations in the diffuse scattering (Fig.~\ref{fig:Diffuse_2time}(a)).
The ``checkerboard'' pattern evident in the two-time correlations during growth reveals a surprising result -- even though the island arrangement is determined by a statistical nucleation process to form each atomic layer of the crystal, the arrangement is correlated across several successive layers, indicating a memory effect in the nucleation locations.
This is similar to the return point memory effect found in magnetic thin films \cite{PiercePRL2003, chesnel2011oscillating, chesnel2013field, chesnel2016shaping,pierce2005disorder}, in which the same pattern of domains can form again after being erased during field cycling.
These changing correlations are clearly produced by the growth process, since the two-time correlations for the periods before and after growth indicate that the island arrangement remains fixed.

\section*{KMC Simulations}

\begin{figure}
\includegraphics[width=0.5\columnwidth]{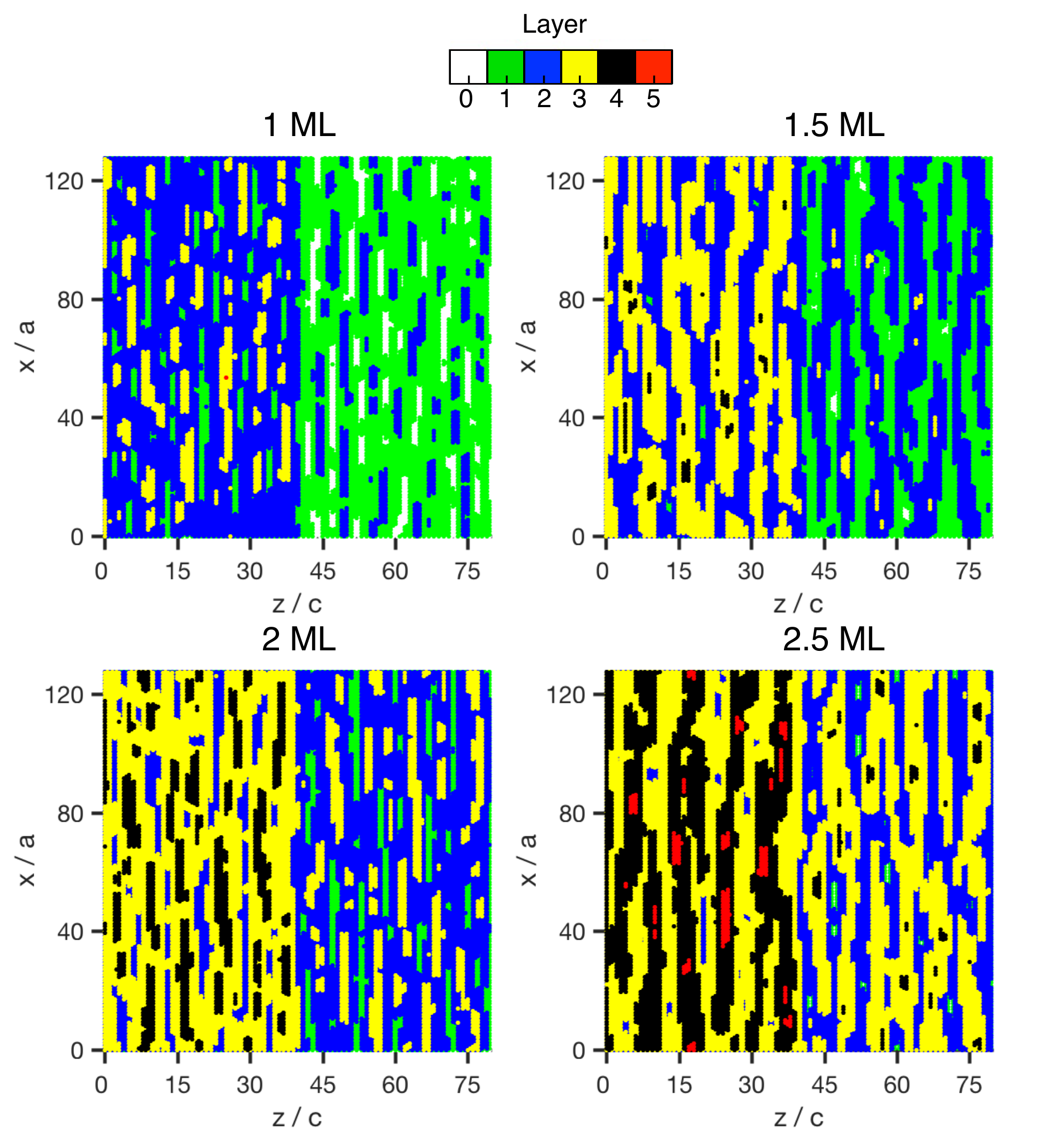}
\caption{\textbf{Correlations of island positions.} Kinetic Monte Carlo simulations of surface island and step arrangements after 1.0, 1.5, 2.0, and 2.5 ML of layer-by-layer growth on m-plane GaN, for one of the 16 random initial conditions used. Colors indicate the surface height, and positions $x$ or $z$ are specified in terms of the number of $a$ or $c$ lattice parameters in the 
{$[1 \, \overline{2} \, 1 \, 0]$}
or
{$[0 \, 0 \, 0 \, 1]$}
directions, respectively. The vicinal surface has an initial step spacing producing terraces half the $z$ width of the simulation. Monolayer-height islands nucleate and coalesce on the terraces to form each layer during growth, and their arrangement is correlated from layer to layer.}
\label{fig:KMC}
\end{figure}

As a comparison to the XPCS experiments, we carried out simulations of layer-by-layer growth on an m-plane GaN surface and calculated the two-time correlations in the simulated diffuse scattering.
Because the simulations are free of crystal defects, this allows us to investigate whether persistence in the island arrangement is intrinsic to the layer-by-layer growth process, or is instead produced by preferred nucleation at fixed defects (as is found for magnetic domains \cite{pierce2005disorder,2018_Martin-Garcia_SciRep8_5991}). 
The KMC model used has been found to reproduce the phenomenology of MOVPE growth on m-plane GaN \cite{xu2017kinetic}.
While the simulation is necessarily carried out on a smaller system than the experiments, the temperature and growth rate in the simulation were chosen to give a similar number of islands per terrace $(\sim 5)$ as in the experiments (Supplementary Section 4).
Figure \ref{fig:Diffuse_2time}(b) shows the simulated two-time correlations.
(For the simulations, the mean intensity $\overline{I}$ in equation~(\ref{eq:ttc}) is calculated by averaging over 16 random initial conditions, rather than by smoothing in $\textbf{Q}$.) 
These show a strong checkerboard pattern from persistence of the island arrangement from layer to layer, even in the absence of crystal defects.
To understand the meaning of the statistical analysis,
it is helpful to examine images of the 2D islands and steps from the simulations after different amounts of growth (Fig.~\ref{fig:KMC}).
One can see that the island arrangement at 2.5 ML has many features in common with that at 1.5 ML.
For example, the yellow regions on the right half of the 2.5 ML image are similar to the blue regions on the right half of the 1.5 ML image.

\section*{Island Arrangement Persistence}

To quantify the island arrangement persistence, we define the quantity $P$ by averaging the two-time correlation over lines running parallel with the main diagonal, representing equal time differences $\Delta t$ perpendicular to the diagonal, 
\begin{equation}
P(\Delta t)
    = \frac{ \left  \langle  
        C(t,t + \Delta t)
            \right \rangle_t}
            { \left  \langle  
        C(t,t + \delta))
            \right \rangle_t}.
\label{eq:corr}
\end{equation}
The numerator is equal to the standard intensity autocorrelation function $g_2 - 1$ \cite{sinha2014x}.
We have normalized $P$ to the value at the minimum non-zero $\Delta t \equiv \delta$, leaving out the point at $\Delta t = 0$ because it contains self-correlated shot noise (Supplementary Section 3).
Both the measured and simulated $P$ values exhibit decaying oscillations (Figs.~\ref{fig:Diffuse_2time}(c) and (d)), with peaks near integer ML positions, indicating a tendency for the island arrangement at any time to repeat after one or more integer monolayers for growth.
In order to extract the main features from this behavior, we have fit $P(\Delta t)$ to the empirical form
\begin{equation}
\begin{split}
 P(\Delta t)& = P_0 + A_{0} \exp(-\Delta t/ \tau_{0})\\ 
 &+ A^{\ast} \exp(-\Delta t/ \tau^{\ast}) \cos(2\pi (\Delta t+\Delta t_0)), 
\end{split}
\label{eq:fit}
\end{equation}
where the 2nd and 3rd terms represent average and oscillating components that decay with time constants $\tau_0$ and $\tau^{\ast}$, respectively.
For display, we have converted the time differences $\Delta t$ and corresponding fit parameters to ML of growth.
This empirical form adequately captures the decay of the oscillations of $P(\Delta t)$ in the measurements, and in the simulations beginning at the first 0.2 ML of growth.

Similar oscillating $g_2$ functions have been observed recently in grazing-incidence small-angle x-ray scattering from surface roughness during sputter deposition of amorphous films \cite{ulbrandt2016direct}.
In that study, the oscillations arise from heterodyne interference between scattering from the surface and from defects in the bulk of the film, and the period of the heterodyne oscillations is inversely related to the product of the wavevector and the velocity vector of the advancing surface.
In our case, the heterodyne period would be $H^{-1} \approx 2.08$~ML per oscillation at our relatively high wavenumber of $H \approx 0.48$~ r.l.u.
We instead see a period of 1~ML per oscillation, equal to the period of the island nucleation and coalescence during layer-by-layer growth.
Thus we are observing homodyne oscillations arising from correlations between island arrangements on subsequent layers, rather than heterodyne effects from the surface velocity. 

\begin{figure}
\includegraphics[width=0.35\linewidth]{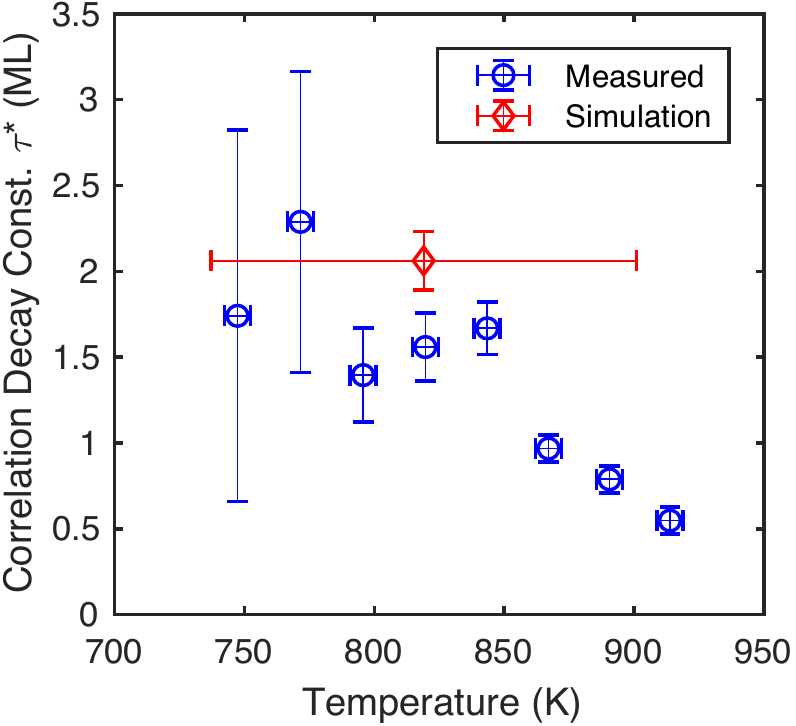}
\caption{\textbf{Persistence times of island arrangements.} Decay constant $\tau^{\ast}$ of the oscillatory part of the correlation functions shown in Fig. \ref{fig:Diffuse_2time} as function of growth temperature.
Measured correlations persist for about 2 ML of growth up to 850~K, in agreement with the simulation.}
\label{fig:correlation_decay}
\end{figure}

We performed the same two-time analysis for eight experimental datasets obtained at growth temperatures $T$ spanning the range across the layer-by-layer growth regime, at fixed growth rate (Supplementary Figs.~6, 7 and Table 2).
The decay constant $\tau^{\ast}$ of the oscillatory part of $P(\Delta t)$ shows a systematic dependence on $T$ (Fig.~\ref{fig:correlation_decay}), with a value of about 2 ML at lower $T$, decreasing at higher $T$ as the transition to step-flow growth at $T \approx 930$~K is approached.
The value of $\tau^{\ast}$ obtained from the simulations agrees reasonably well with the experiments, although correspondence between the simulation and experimental temperatures is uncertain (Supplementary Section 4).

\begin{figure}
\includegraphics[trim=50 200 50 100,clip,width=0.8\linewidth]{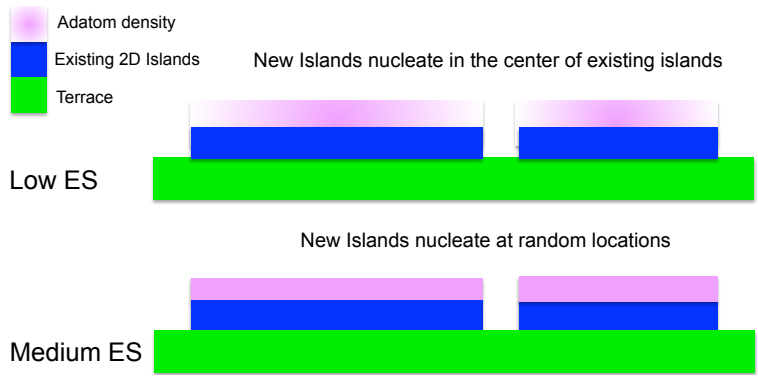}
\caption{\textbf{Mechanism for island arrangement persistence.} Schematic atomic-scale cross sections during layer-by-layer growth, showing dilute adatom density (pink) on tops of existing 2D islands just before they coalesce (blue), for systems with low and medium Ehrlich-Schwoebel (ES) barriers. 
For low ES barriers, the island edges act as a sink for adatoms, so their density (and thus the nucleation rate of islands in the next layer) is maximum in the center of islands. For medium ES barriers, adatom densities are more uniform, leading to more random nucleation locations.
High ES barriers (not shown) produce 3D growth.}
\label{fig:ES}
\end{figure}

Based on the physics underlying island nucleation, we propose the following mechanism for the observed island arrangement persistence in layer-by-layer growth.
As one layer completes, nucleation events for 2D islands in the next layer preferentially occur in regions of high adatom density.
The adatom density distribution is affected by the location of island and step edges in the almost-complete layer.
Analysis of step-flow growth indicates that adatom distributions above a step edge depend upon the magnitude of the ES barrier \cite{kaufmann2016critical}, and the same phenomenon will affect adatom distributions on top of existing islands during layer-by-layer growth.
When the ES barrier is low, the island edges act as sinks for adatoms, so the maximum adatom density will be in the center of the islands (Fig.~\ref{fig:ES}).
This will favor nucleation of islands in the next layer to occur near the nucleation locations of the previous layer, leading to persistent island arrangements.
If the ES barrier has a medium value, the edges are less effective sinks, the adatom density and nucleation locations are more uniform, and the persistence will be lower.
A high ES barrier will inhibit island coalescence, changing the growth mode from layer-by-layer to 3D.
Thus island arrangement persistence revealed by XPCS provides a sensitive new probe of interlayer transport during crystal growth.

In summary, we have performed the first XPCS measurements of 2D island dynamics during layer by layer growth.
The experiments demonstrate that the full bandwidth of an undulator harmonic can be used for XPCS studies of atomic-scale surface dynamics, making it possible to obtain sufficient signal to measure two-time correlations in the diffuse scattering from monolayer-height islands. 
The observed correlations show a new phenomenon, persistent 2D island arrangements during layer-by-layer growth. 
The phenomenon is also observed in kinetic Monte Carlo simulations of layer-by layer growth, indicating that it arises by communication of the arrangement of islands in the underlying layer to those in the newly forming layer via the adatom density distribution.
The XPCS results provide new ways to test models of crystal growth and to understand the atomic scale phenomena that allow us to control and optimize materials synthesis processes.
For example, atomic step patterns on substrate surfaces can directly control the domain patterns in deposited epitaxial layers \cite{2008_Thompson_APL93_182901,2016_Mattoni_NatCom7_13141}.

The two-time correlations discussed here are but one of many higher-order correlation functions that can be explored by coherent x-ray methods \cite{ravy2013homometry,shpyrko2014x},
providing a richer view of the atomic-scale processes that give rise to ordered nanostructures.
The penetrating nature of high energy x-rays will allow these studies in native operating environments critical to applications.
The methods developed here for the model GaN MOVPE growth system open the way for studies of surface dynamics in many other areas, such as growth of bulk crystals or quasi-two-dimensional materials in reactive sputtering or pulsed laser deposition, or from the liquid phase.
The greatly enhanced coherence of x-ray beams soon to be provided by new and upgraded hard x-ray sources \cite{hettel2014dlsr} will produce speckle in the scattering from every system, enabling XPCS studies across broad areas of physics.

\section*{Methods}

The experiments were carried out at beamline 12ID-D at the Advanced Photon Source (APS), using a goniometer and MOVPE system specifically developed for coherent x-ray studies \cite{ju2017instrument}.
An intense ``pink beam'' with appropriate coherence characteristics was prepared using a setup similar to that described previously \cite{ju2018characterization}, with the addition of a harmonic rejection mirror pair (Supplementary Section 1).
The beam incident on the sample had a typical intensity of $1.4 \times 10^{12}$ photons per second at $E = 25.75$ keV, in a spot size of $4 \times 16$ $\mu$m with transverse coherence lengths of $1.2 \times 2.7$ $\mu$m
in the vertical and horizontal, respectively.
X-ray speckle patterns were recorded using a photon counting area detector with a GaAs sensor having 512 $\times$ 512 pixels, 55 $\mu$m pixel size, located 2.39 m from the sample (Amsterdam Scientific Instruments LynX 1800). 
Data were corrected for the measured flat-field response of the detector.
Dynamics were recorded with a sequence of 10~s integrations, subsequently binned by 3 in time. 

MOVPE growth of GaN on the m-plane surface was carried out using procedures and substrates described previously \cite{perret2017island}.
Substrate temperatures were calibrated to within $\pm5$~K \cite{ju2017instrument}.
Growth was started or stopped by injecting or venting a supply of the Ga precursor triethylgallium (TEGa), with a large oversupply of the N precursor (NH$_3$) constantly supplied.
Growth runs were performed at eight different temperatures using the same
TEGa flow ($4.0 \times 10^{-3}$ $\mu$mole/min).
The same substrate was used for all growths; it was annealed for 300~s at $T = 1120$~K between growths to recover the same initial surface condition with minimal islands.
To account for thermal expansion of the heater, the position of the sample was realigned between growths, so slightly different regions of the surface were investigated at each temperature.
Growth rates were extracted from the oscillation periods of the diffuse and CTR scattering. 
They varied by about 10\% from growth to growth around a mean of $1.25 \times 10^{-3}$ ML/s (Supplementary Table 2).
The growth rate was constant during a given growth, apart from a $\sim5\%$ lower growth rate for the first ML.

Calculated two-time correlations in the scattering from islands were derived from KMC simulations on the m-plane surface of a three-dimensional HCP lattice \cite{xu2017kinetic}. 
The dynamics of deposition and diffusion on a stepped surface were simulated using a variable time step method as implemented in the Stochastic Parallel PARticle Kinetic Simulator (SPPARKS) computer code\cite{2009_Plimpton_SAND2009-6226}
(Supplementary Section 4). 

Raw x-ray data were generated at the Advanced Photon Source large-scale facility. Derived data and simulations supporting the findings of this study are available from the corresponding author upon request.

\bibliography{2018_GJu_TwoTime_m_plane_XPCS}

\begin{thebibliography}{44}%
\makeatletter
\providecommand \@ifxundefined [1]{%
 \@ifx{#1\undefined}
}%
\providecommand \@ifnum [1]{%
 \ifnum #1\expandafter \@firstoftwo
 \else \expandafter \@secondoftwo
 \fi
}%
\providecommand \@ifx [1]{%
 \ifx #1\expandafter \@firstoftwo
 \else \expandafter \@secondoftwo
 \fi
}%
\providecommand \natexlab [1]{#1}%
\providecommand \enquote  [1]{``#1''}%
\providecommand \bibnamefont  [1]{#1}%
\providecommand \bibfnamefont [1]{#1}%
\providecommand \citenamefont [1]{#1}%
\providecommand \href@noop [0]{\@secondoftwo}%
\providecommand \href [0]{\begingroup \@sanitize@url \@href}%
\providecommand \@href[1]{\@@startlink{#1}\@@href}%
\providecommand \@@href[1]{\endgroup#1\@@endlink}%
\providecommand \@sanitize@url [0]{\catcode `\\12\catcode `\$12\catcode
  `\&12\catcode `\#12\catcode `\^12\catcode `\_12\catcode `\%12\relax}%
\providecommand \@@startlink[1]{}%
\providecommand \@@endlink[0]{}%
\providecommand \url  [0]{\begingroup\@sanitize@url \@url }%
\providecommand \@url [1]{\endgroup\@href {#1}{\urlprefix }}%
\providecommand \urlprefix  [0]{URL }%
\providecommand \Eprint [0]{\href }%
\providecommand \doibase [0]{http://dx.doi.org/}%
\providecommand \selectlanguage [0]{\@gobble}%
\providecommand \bibinfo  [0]{\@secondoftwo}%
\providecommand \bibfield  [0]{\@secondoftwo}%
\providecommand \translation [1]{[#1]}%
\providecommand \BibitemOpen [0]{}%
\providecommand \bibitemStop [0]{}%
\providecommand \bibitemNoStop [0]{.\EOS\space}%
\providecommand \EOS [0]{\spacefactor3000\relax}%
\providecommand \BibitemShut  [1]{\csname bibitem#1\endcsname}%
\let\auto@bib@innerbib\@empty
\bibitem [{\citenamefont {Song}\ \emph {et~al.}(2013)\citenamefont {Song},
  \citenamefont {Chen}, \citenamefont {Dabade}, \citenamefont {Shield},\ and\
  \citenamefont {James}}]{2013_Song_Nature502_85}%
  \BibitemOpen
  \bibfield  {author} {\bibinfo {author} {\bibfnamefont {Yintao}\ \bibnamefont
  {Song}}, \bibinfo {author} {\bibfnamefont {Xian}\ \bibnamefont {Chen}},
  \bibinfo {author} {\bibfnamefont {Vivekanand}\ \bibnamefont {Dabade}},
  \bibinfo {author} {\bibfnamefont {Thomas~W.}\ \bibnamefont {Shield}}, \ and\
  \bibinfo {author} {\bibfnamefont {Richard~D.}\ \bibnamefont {James}},\
  }\bibfield  {title} {\enquote {\bibinfo {title} {Enhanced reversibility and
  unusual microstructure of a phase-transforming material},}\ }\href {\doibase
  10.1038/nature12532} {\bibfield  {journal} {\bibinfo  {journal} {Nature}\
  }\textbf {\bibinfo {volume} {502}},\ \bibinfo {pages} {85--88} (\bibinfo
  {year} {2013})}\BibitemShut {NoStop}%
\bibitem [{\citenamefont {Ogawa}\ \emph {et~al.}(2016)\citenamefont {Ogawa},
  \citenamefont {Ando}, \citenamefont {Sutou},\ and\ \citenamefont
  {Koike}}]{2016_Ogawa_Science353_368}%
  \BibitemOpen
  \bibfield  {author} {\bibinfo {author} {\bibfnamefont {Yukiko}\ \bibnamefont
  {Ogawa}}, \bibinfo {author} {\bibfnamefont {Daisuke}\ \bibnamefont {Ando}},
  \bibinfo {author} {\bibfnamefont {Yuji}\ \bibnamefont {Sutou}}, \ and\
  \bibinfo {author} {\bibfnamefont {Junichi}\ \bibnamefont {Koike}},\
  }\bibfield  {title} {\enquote {\bibinfo {title} {A lightweight shape-memory
  magnesium alloy},}\ }\href {\doibase 10.1126/science.aaf6524} {\bibfield
  {journal} {\bibinfo  {journal} {Science}\ }\textbf {\bibinfo {volume}
  {353}},\ \bibinfo {pages} {368--370} (\bibinfo {year} {2016})},\ \Eprint
  {http://arxiv.org/abs/http://science.sciencemag.org/content/353/6297/368.full.pdf}
  {http://science.sciencemag.org/content/353/6297/368.full.pdf} \BibitemShut
  {NoStop}%
\bibitem [{\citenamefont {Gornostyrev}\ and\ \citenamefont
  {Katsnelson}(2015)}]{2015_Gornostyref_PCCP17_27249}%
  \BibitemOpen
  \bibfield  {author} {\bibinfo {author} {\bibfnamefont {Yu.~N.}\ \bibnamefont
  {Gornostyrev}}\ and\ \bibinfo {author} {\bibfnamefont {M.~I.}\ \bibnamefont
  {Katsnelson}},\ }\bibfield  {title} {\enquote {\bibinfo {title} {Misfit
  stabilized embedded nanoparticles in metallic alloys},}\ }\href {\doibase
  10.1039/C5CP04641F} {\bibfield  {journal} {\bibinfo  {journal} {Physical
  Chemistry Chemical Physics}\ }\textbf {\bibinfo {volume} {17}},\ \bibinfo
  {pages} {27249--27257} (\bibinfo {year} {2015})}\BibitemShut {NoStop}%
\bibitem [{\citenamefont {Mikhailov}\ and\ \citenamefont
  {Showalter}(2006)}]{2006_Mikhailov_PhysRep425_79}%
  \BibitemOpen
  \bibfield  {author} {\bibinfo {author} {\bibfnamefont {Alexander~S.}\
  \bibnamefont {Mikhailov}}\ and\ \bibinfo {author} {\bibfnamefont {Kenneth}\
  \bibnamefont {Showalter}},\ }\bibfield  {title} {\enquote {\bibinfo {title}
  {Control of waves, patterns and turbulence in chemical systems},}\ }\href
  {\doibase 10.1016/j.physrep.2005.11.003} {\bibfield  {journal} {\bibinfo
  {journal} {Physics Reports}\ }\textbf {\bibinfo {volume} {425}},\ \bibinfo
  {pages} {79--194} (\bibinfo {year} {2006})}\BibitemShut {NoStop}%
\bibitem [{\citenamefont {Jiang}\ \emph {et~al.}(2008)\citenamefont {Jiang},
  \citenamefont {Munkholm}, \citenamefont {Wang}, \citenamefont {Streiffer},
  \citenamefont {Thompson}, \citenamefont {Fuoss}, \citenamefont {Latifi},
  \citenamefont {Elder},\ and\ \citenamefont
  {Stephenson}}]{2008_Jiang_PRL101_086102}%
  \BibitemOpen
  \bibfield  {author} {\bibinfo {author} {\bibfnamefont {Fan}\ \bibnamefont
  {Jiang}}, \bibinfo {author} {\bibfnamefont {A.}~\bibnamefont {Munkholm}},
  \bibinfo {author} {\bibfnamefont {R.~V}\ \bibnamefont {Wang}}, \bibinfo
  {author} {\bibfnamefont {S.~K.}\ \bibnamefont {Streiffer}}, \bibinfo {author}
  {\bibfnamefont {Carol}\ \bibnamefont {Thompson}}, \bibinfo {author}
  {\bibfnamefont {P.~H.}\ \bibnamefont {Fuoss}}, \bibinfo {author}
  {\bibfnamefont {K.}~\bibnamefont {Latifi}}, \bibinfo {author} {\bibfnamefont
  {K.~R.}\ \bibnamefont {Elder}}, \ and\ \bibinfo {author} {\bibfnamefont
  {G.~B.}\ \bibnamefont {Stephenson}},\ }\bibfield  {title} {\enquote {\bibinfo
  {title} {Spontaneous oscillations and waves during chemical vapor deposition
  of {InN}},}\ }\href {\doibase 10.1103/PhysRevLett.101.086102} {\bibfield
  {journal} {\bibinfo  {journal} {Physical Review Letters}\ }\textbf {\bibinfo
  {volume} {101}},\ \bibinfo {pages} {086102} (\bibinfo {year}
  {2008})}\BibitemShut {NoStop}%
\bibitem [{\citenamefont {McLeod}\ \emph {et~al.}(2016)\citenamefont {McLeod},
  \citenamefont {van Heumen}, \citenamefont {Ramirez}, \citenamefont {Wang},
  \citenamefont {Saerbeck}, \citenamefont {Guenon}, \citenamefont {Goldflam},
  \citenamefont {Anderegg}, \citenamefont {Kelly}, \citenamefont {Mueller},
  \citenamefont {Liu}, \citenamefont {Schuller},\ and\ \citenamefont
  {Basov}}]{2016_McLeod_NatPhys13_80}%
  \BibitemOpen
  \bibfield  {author} {\bibinfo {author} {\bibfnamefont {A.~S.}\ \bibnamefont
  {McLeod}}, \bibinfo {author} {\bibfnamefont {E.}~\bibnamefont {van Heumen}},
  \bibinfo {author} {\bibfnamefont {J.~G.}\ \bibnamefont {Ramirez}}, \bibinfo
  {author} {\bibfnamefont {S.}~\bibnamefont {Wang}}, \bibinfo {author}
  {\bibfnamefont {T.}~\bibnamefont {Saerbeck}}, \bibinfo {author}
  {\bibfnamefont {S.}~\bibnamefont {Guenon}}, \bibinfo {author} {\bibfnamefont
  {M.}~\bibnamefont {Goldflam}}, \bibinfo {author} {\bibfnamefont
  {L.}~\bibnamefont {Anderegg}}, \bibinfo {author} {\bibfnamefont
  {P.}~\bibnamefont {Kelly}}, \bibinfo {author} {\bibfnamefont
  {A.}~\bibnamefont {Mueller}}, \bibinfo {author} {\bibfnamefont {M.~K.}\
  \bibnamefont {Liu}}, \bibinfo {author} {\bibfnamefont {Ivan~K.}\ \bibnamefont
  {Schuller}}, \ and\ \bibinfo {author} {\bibfnamefont {D.~N.}\ \bibnamefont
  {Basov}},\ }\bibfield  {title} {\enquote {\bibinfo {title} {Nanotextured
  phase coexistence in the correlated insulator {V$_2$O$_3$}},}\ }\href
  {\doibase 10.1038/nphys3882} {\bibfield  {journal} {\bibinfo  {journal}
  {Nature Physics}\ }\textbf {\bibinfo {volume} {13}},\ \bibinfo {pages}
  {80--86} (\bibinfo {year} {2016})}\BibitemShut {NoStop}%
\bibitem [{\citenamefont {Huang}\ and\ \citenamefont
  {Cheong}(2017)}]{2017_Huang_NatRevMat2_17004}%
  \BibitemOpen
  \bibfield  {author} {\bibinfo {author} {\bibfnamefont {Fei-Ting}\
  \bibnamefont {Huang}}\ and\ \bibinfo {author} {\bibfnamefont {Sang-Wook}\
  \bibnamefont {Cheong}},\ }\bibfield  {title} {\enquote {\bibinfo {title}
  {Aperiodic topological order in the domain configurations of functional
  materials},}\ }\href {\doibase 10.1038/natrevmats.2017.4} {\bibfield
  {journal} {\bibinfo  {journal} {Nature Reviews Materials}\ }\textbf {\bibinfo
  {volume} {2}},\ \bibinfo {pages} {17004} (\bibinfo {year}
  {2017})}\BibitemShut {NoStop}%
\bibitem [{\citenamefont {Shpyrko}(2014)}]{shpyrko2014x}%
  \BibitemOpen
  \bibfield  {author} {\bibinfo {author} {\bibfnamefont {Oleg~G.}\ \bibnamefont
  {Shpyrko}},\ }\bibfield  {title} {\enquote {\bibinfo {title} {X-ray photon
  correlation spectroscopy},}\ }\href {\doibase 10.1107/S1600577514018232}
  {\bibfield  {journal} {\bibinfo  {journal} {Journal of Synchrotron
  Radiation}\ }\textbf {\bibinfo {volume} {21}},\ \bibinfo {pages} {1057--1064}
  (\bibinfo {year} {2014})}\BibitemShut {NoStop}%
\bibitem [{\citenamefont {Pierce}\ \emph {et~al.}(2009)\citenamefont {Pierce},
  \citenamefont {Chang}, \citenamefont {Hennessy}, \citenamefont {Komanicky},
  \citenamefont {Sprung}, \citenamefont {Sandy},\ and\ \citenamefont
  {You}}]{pierce2009surface}%
  \BibitemOpen
  \bibfield  {author} {\bibinfo {author} {\bibfnamefont {M.~S.}\ \bibnamefont
  {Pierce}}, \bibinfo {author} {\bibfnamefont {K.~C.}\ \bibnamefont {Chang}},
  \bibinfo {author} {\bibfnamefont {D.}~\bibnamefont {Hennessy}}, \bibinfo
  {author} {\bibfnamefont {V.}~\bibnamefont {Komanicky}}, \bibinfo {author}
  {\bibfnamefont {M.}~\bibnamefont {Sprung}}, \bibinfo {author} {\bibfnamefont
  {A.}~\bibnamefont {Sandy}}, \ and\ \bibinfo {author} {\bibfnamefont
  {H.}~\bibnamefont {You}},\ }\bibfield  {title} {\enquote {\bibinfo {title}
  {Surface x-ray speckles: coherent surface diffraction from {Au} {(001)}},}\
  }\href {\doibase 10.1103/PhysRevLett.103.165501} {\bibfield  {journal}
  {\bibinfo  {journal} {Physical Review Letters}\ }\textbf {\bibinfo {volume}
  {103}},\ \bibinfo {pages} {165501} (\bibinfo {year} {2009})}\BibitemShut
  {NoStop}%
\bibitem [{\citenamefont {Kim}\ \emph {et~al.}(2003)\citenamefont {Kim},
  \citenamefont {R{\"u}hm}, \citenamefont {Lurio}, \citenamefont {Basu},
  \citenamefont {Lal}, \citenamefont {Lumma}, \citenamefont {Mochrie},\ and\
  \citenamefont {Sinha}}]{kim2003surface}%
  \BibitemOpen
  \bibfield  {author} {\bibinfo {author} {\bibfnamefont {Hyunjung}\
  \bibnamefont {Kim}}, \bibinfo {author} {\bibfnamefont {A.}~\bibnamefont
  {R{\"u}hm}}, \bibinfo {author} {\bibfnamefont {L.~B.}\ \bibnamefont {Lurio}},
  \bibinfo {author} {\bibfnamefont {J.~K.}\ \bibnamefont {Basu}}, \bibinfo
  {author} {\bibfnamefont {J.}~\bibnamefont {Lal}}, \bibinfo {author}
  {\bibfnamefont {D.}~\bibnamefont {Lumma}}, \bibinfo {author} {\bibfnamefont
  {S.~G.~J.}\ \bibnamefont {Mochrie}}, \ and\ \bibinfo {author} {\bibfnamefont
  {S.~K.}\ \bibnamefont {Sinha}},\ }\bibfield  {title} {\enquote {\bibinfo
  {title} {Surface dynamics of polymer films},}\ }\href {\doibase
  10.1103/PhysRevLett.90.068302} {\bibfield  {journal} {\bibinfo  {journal}
  {Physical Review Letters}\ }\textbf {\bibinfo {volume} {90}},\ \bibinfo
  {pages} {068302} (\bibinfo {year} {2003})}\BibitemShut {NoStop}%
\bibitem [{\citenamefont {Ruta}\ \emph {et~al.}(2014)\citenamefont {Ruta},
  \citenamefont {Baldi}, \citenamefont {Chushkin}, \citenamefont {Ruffl{\'e}},
  \citenamefont {Cristofolini}, \citenamefont {Fontana}, \citenamefont
  {Zanatta},\ and\ \citenamefont {Nazzani}}]{ruta2014revealing}%
  \BibitemOpen
  \bibfield  {author} {\bibinfo {author} {\bibfnamefont {B.}~\bibnamefont
  {Ruta}}, \bibinfo {author} {\bibfnamefont {G.}~\bibnamefont {Baldi}},
  \bibinfo {author} {\bibfnamefont {Y.}~\bibnamefont {Chushkin}}, \bibinfo
  {author} {\bibfnamefont {Benoit}\ \bibnamefont {Ruffl{\'e}}}, \bibinfo
  {author} {\bibfnamefont {L.}~\bibnamefont {Cristofolini}}, \bibinfo {author}
  {\bibfnamefont {Adriano}\ \bibnamefont {Fontana}}, \bibinfo {author}
  {\bibfnamefont {M.}~\bibnamefont {Zanatta}}, \ and\ \bibinfo {author}
  {\bibfnamefont {Francesco}\ \bibnamefont {Nazzani}},\ }\bibfield  {title}
  {\enquote {\bibinfo {title} {Revealing the fast atomic motion of network
  glasses},}\ }\href {\doibase 10.1038/ncomms4939} {\bibfield  {journal}
  {\bibinfo  {journal} {Nature Communications}\ }\textbf {\bibinfo {volume}
  {5}},\ \bibinfo {pages} {3939} (\bibinfo {year} {2014})}\BibitemShut
  {NoStop}%
\bibitem [{\citenamefont {Roseker}\ \emph {et~al.}(2018)\citenamefont
  {Roseker}, \citenamefont {Hruszkewycz}, \citenamefont {Lehmk{\"u}hler},
  \citenamefont {Walther}, \citenamefont {Schulte-Schrepping}, \citenamefont
  {Lee}, \citenamefont {Osaka}, \citenamefont {Str{\"u}der}, \citenamefont
  {Hartmann}, \citenamefont {Sikorski}, \citenamefont {Song}, \citenamefont
  {Robert}, \citenamefont {Fuoss}, \citenamefont {Sutton}, \citenamefont
  {Stephenson},\ and\ \citenamefont {Gr{\"u}bel}}]{roseker2018NatComm}%
  \BibitemOpen
  \bibfield  {author} {\bibinfo {author} {\bibfnamefont {W.}~\bibnamefont
  {Roseker}}, \bibinfo {author} {\bibfnamefont {S.~O.}\ \bibnamefont
  {Hruszkewycz}}, \bibinfo {author} {\bibfnamefont {F.}~\bibnamefont
  {Lehmk{\"u}hler}}, \bibinfo {author} {\bibfnamefont {M.}~\bibnamefont
  {Walther}}, \bibinfo {author} {\bibfnamefont {H.}~\bibnamefont
  {Schulte-Schrepping}}, \bibinfo {author} {\bibfnamefont {S.}~\bibnamefont
  {Lee}}, \bibinfo {author} {\bibfnamefont {T.}~\bibnamefont {Osaka}}, \bibinfo
  {author} {\bibfnamefont {L}~\bibnamefont {Str{\"u}der}}, \bibinfo {author}
  {\bibfnamefont {R.}~\bibnamefont {Hartmann}}, \bibinfo {author}
  {\bibfnamefont {M.}~\bibnamefont {Sikorski}}, \bibinfo {author}
  {\bibfnamefont {S.}~\bibnamefont {Song}}, \bibinfo {author} {\bibfnamefont
  {A.}~\bibnamefont {Robert}}, \bibinfo {author} {\bibfnamefont {P.~H.}\
  \bibnamefont {Fuoss}}, \bibinfo {author} {\bibfnamefont {M.}~\bibnamefont
  {Sutton}}, \bibinfo {author} {\bibfnamefont {G.~B.}\ \bibnamefont
  {Stephenson}}, \ and\ \bibinfo {author} {\bibfnamefont {G.}~\bibnamefont
  {Gr{\"u}bel}},\ }\bibfield  {title} {\enquote {\bibinfo {title} {Towards
  ultrafast dynamics with split-pulse x-ray photon correlation spectroscopy at
  free electron laser sources},}\ }\href@noop {} {\bibfield  {journal}
  {\bibinfo  {journal} {Nature Communications}\ } (\bibinfo {year} {2018})},\
  \bibinfo {note} {in press}\BibitemShut {NoStop}%
\bibitem [{\citenamefont {Ulbrandt}\ \emph {et~al.}(2016)\citenamefont
  {Ulbrandt}, \citenamefont {Rainville}, \citenamefont {Wagenbach},
  \citenamefont {Narayanan}, \citenamefont {Sandy}, \citenamefont {Zhou},
  \citenamefont {Ludwig~Jr.},\ and\ \citenamefont
  {Headrick}}]{ulbrandt2016direct}%
  \BibitemOpen
  \bibfield  {author} {\bibinfo {author} {\bibfnamefont {Jeffrey~G.}\
  \bibnamefont {Ulbrandt}}, \bibinfo {author} {\bibfnamefont {Meliha~G.}\
  \bibnamefont {Rainville}}, \bibinfo {author} {\bibfnamefont {Christa}\
  \bibnamefont {Wagenbach}}, \bibinfo {author} {\bibfnamefont {Suresh}\
  \bibnamefont {Narayanan}}, \bibinfo {author} {\bibfnamefont {Alec~R.}\
  \bibnamefont {Sandy}}, \bibinfo {author} {\bibfnamefont {Hua}\ \bibnamefont
  {Zhou}}, \bibinfo {author} {\bibfnamefont {Karl~F.}\ \bibnamefont
  {Ludwig~Jr.}}, \ and\ \bibinfo {author} {\bibfnamefont {Randall~L.}\
  \bibnamefont {Headrick}},\ }\bibfield  {title} {\enquote {\bibinfo {title}
  {Direct measurement of the propagation velocity of defects using coherent
  x-rays},}\ }\href {\doibase 10.1038/nphys3708} {\bibfield  {journal}
  {\bibinfo  {journal} {Nature Physics}\ }\textbf {\bibinfo {volume} {12}},\
  \bibinfo {pages} {794--799} (\bibinfo {year} {2016})}\BibitemShut {NoStop}%
\bibitem [{\citenamefont {Pierce}\ \emph {et~al.}(2003)\citenamefont {Pierce},
  \citenamefont {Moore}, \citenamefont {Sorensen}, \citenamefont {Kevan},
  \citenamefont {Hellwig}, \citenamefont {Fullerton},\ and\ \citenamefont
  {Kortright}}]{PiercePRL2003}%
  \BibitemOpen
  \bibfield  {author} {\bibinfo {author} {\bibfnamefont {Michael~S.}\
  \bibnamefont {Pierce}}, \bibinfo {author} {\bibfnamefont {Rob~G.}\
  \bibnamefont {Moore}}, \bibinfo {author} {\bibfnamefont {Larry~B.}\
  \bibnamefont {Sorensen}}, \bibinfo {author} {\bibfnamefont {Stephen~D.}\
  \bibnamefont {Kevan}}, \bibinfo {author} {\bibfnamefont {Olav}\ \bibnamefont
  {Hellwig}}, \bibinfo {author} {\bibfnamefont {Eric~E.}\ \bibnamefont
  {Fullerton}}, \ and\ \bibinfo {author} {\bibfnamefont {Jeffrey~B.}\
  \bibnamefont {Kortright}},\ }\bibfield  {title} {\enquote {\bibinfo {title}
  {Quasistatic x-ray speckle metrology of microscopic magnetic return-point
  memory},}\ }\href {\doibase 10.1103/PhysRevLett.90.175502} {\bibfield
  {journal} {\bibinfo  {journal} {Physical Review Letters}\ }\textbf {\bibinfo
  {volume} {90}},\ \bibinfo {pages} {175502} (\bibinfo {year}
  {2003})}\BibitemShut {NoStop}%
\bibitem [{\citenamefont {Sanborn}\ \emph {et~al.}(2011)\citenamefont
  {Sanborn}, \citenamefont {Ludwig}, \citenamefont {Rogers},\ and\
  \citenamefont {Sutton}}]{SanbornPRL2011}%
  \BibitemOpen
  \bibfield  {author} {\bibinfo {author} {\bibfnamefont {Christopher}\
  \bibnamefont {Sanborn}}, \bibinfo {author} {\bibfnamefont {Karl~F.}\
  \bibnamefont {Ludwig}}, \bibinfo {author} {\bibfnamefont {Michael~C.}\
  \bibnamefont {Rogers}}, \ and\ \bibinfo {author} {\bibfnamefont {Mark}\
  \bibnamefont {Sutton}},\ }\bibfield  {title} {\enquote {\bibinfo {title}
  {Direct measurement of microstructural avalanches during the martensitic
  transition of cobalt using coherent x-ray scattering},}\ }\href {\doibase
  10.1103/PhysRevLett.107.015702} {\bibfield  {journal} {\bibinfo  {journal}
  {Physical Review Letters}\ }\textbf {\bibinfo {volume} {107}},\ \bibinfo
  {pages} {015702} (\bibinfo {year} {2011})}\BibitemShut {NoStop}%
\bibitem [{\citenamefont {Chesnel}\ \emph {et~al.}(2011)\citenamefont
  {Chesnel}, \citenamefont {Nelson}, \citenamefont {Kevan}, \citenamefont
  {Carey},\ and\ \citenamefont {Fullerton}}]{chesnel2011oscillating}%
  \BibitemOpen
  \bibfield  {author} {\bibinfo {author} {\bibfnamefont {Karine}\ \bibnamefont
  {Chesnel}}, \bibinfo {author} {\bibfnamefont {Joseph~A.}\ \bibnamefont
  {Nelson}}, \bibinfo {author} {\bibfnamefont {Stephen~D.}\ \bibnamefont
  {Kevan}}, \bibinfo {author} {\bibfnamefont {Matthew~J.}\ \bibnamefont
  {Carey}}, \ and\ \bibinfo {author} {\bibfnamefont {Eric~E.}\ \bibnamefont
  {Fullerton}},\ }\bibfield  {title} {\enquote {\bibinfo {title} {Oscillating
  spatial dependence of domain memory in ferromagnetic films mapped via x-ray
  speckle correlation},}\ }\href {\doibase 10.1103/PhysRevB.83.054436}
  {\bibfield  {journal} {\bibinfo  {journal} {Physical Review B}\ }\textbf
  {\bibinfo {volume} {83}},\ \bibinfo {pages} {054436} (\bibinfo {year}
  {2011})}\BibitemShut {NoStop}%
\bibitem [{\citenamefont {Chesnel}\ \emph {et~al.}(2013)\citenamefont
  {Chesnel}, \citenamefont {Wilcken}, \citenamefont {Rytting}, \citenamefont
  {Kevan},\ and\ \citenamefont {Fullerton}}]{chesnel2013field}%
  \BibitemOpen
  \bibfield  {author} {\bibinfo {author} {\bibfnamefont {Karine}\ \bibnamefont
  {Chesnel}}, \bibinfo {author} {\bibfnamefont {Brian}\ \bibnamefont
  {Wilcken}}, \bibinfo {author} {\bibfnamefont {Matthew}\ \bibnamefont
  {Rytting}}, \bibinfo {author} {\bibfnamefont {Steve~D.}\ \bibnamefont
  {Kevan}}, \ and\ \bibinfo {author} {\bibfnamefont {Eric~E.}\ \bibnamefont
  {Fullerton}},\ }\bibfield  {title} {\enquote {\bibinfo {title} {Field mapping
  and temperature dependence of magnetic domain memory induced by exchange
  couplings},}\ }\href {\doibase 10.1088/1367-2630/15/2/023016} {\bibfield
  {journal} {\bibinfo  {journal} {New Journal of Physics}\ }\textbf {\bibinfo
  {volume} {15}},\ \bibinfo {pages} {023016} (\bibinfo {year}
  {2013})}\BibitemShut {NoStop}%
\bibitem [{\citenamefont {Chesnel}\ \emph {et~al.}(2016)\citenamefont
  {Chesnel}, \citenamefont {Safsten}, \citenamefont {Rytting},\ and\
  \citenamefont {Fullerton}}]{chesnel2016shaping}%
  \BibitemOpen
  \bibfield  {author} {\bibinfo {author} {\bibfnamefont {Karine}\ \bibnamefont
  {Chesnel}}, \bibinfo {author} {\bibfnamefont {Alex}\ \bibnamefont {Safsten}},
  \bibinfo {author} {\bibfnamefont {Matthew}\ \bibnamefont {Rytting}}, \ and\
  \bibinfo {author} {\bibfnamefont {Eric~E.}\ \bibnamefont {Fullerton}},\
  }\bibfield  {title} {\enquote {\bibinfo {title} {Shaping nanoscale magnetic
  domain memory in exchange-coupled ferromagnets by field cooling},}\ }\href
  {\doibase 10.1038/ncomms11648} {\bibfield  {journal} {\bibinfo  {journal}
  {Nature Communications}\ }\textbf {\bibinfo {volume} {7}},\ \bibinfo {pages}
  {11648} (\bibinfo {year} {2016})}\BibitemShut {NoStop}%
\bibitem [{\citenamefont {Pierce}\ \emph {et~al.}(2005)\citenamefont {Pierce},
  \citenamefont {Buechler}, \citenamefont {Sorensen}, \citenamefont {Turner},
  \citenamefont {Kevan}, \citenamefont {Jagla}, \citenamefont {Deutsch},
  \citenamefont {Mai}, \citenamefont {Narayan}, \citenamefont {Davies},
  \citenamefont {Liu}, \citenamefont {Dunn}, \citenamefont {Chesnel},
  \citenamefont {Kortright}, \citenamefont {Hellwig},\ and\ \citenamefont
  {Fullerton.}}]{pierce2005disorder}%
  \BibitemOpen
  \bibfield  {author} {\bibinfo {author} {\bibfnamefont {M.~S.}\ \bibnamefont
  {Pierce}}, \bibinfo {author} {\bibfnamefont {C.~R.}\ \bibnamefont
  {Buechler}}, \bibinfo {author} {\bibfnamefont {L.~B.}\ \bibnamefont
  {Sorensen}}, \bibinfo {author} {\bibfnamefont {J.~J.}\ \bibnamefont
  {Turner}}, \bibinfo {author} {\bibfnamefont {S.~D.}\ \bibnamefont {Kevan}},
  \bibinfo {author} {\bibfnamefont {E.~A.}\ \bibnamefont {Jagla}}, \bibinfo
  {author} {\bibfnamefont {J.~M.}\ \bibnamefont {Deutsch}}, \bibinfo {author}
  {\bibfnamefont {T.}~\bibnamefont {Mai}}, \bibinfo {author} {\bibfnamefont
  {O.}~\bibnamefont {Narayan}}, \bibinfo {author} {\bibfnamefont {J.~E.}\
  \bibnamefont {Davies}}, \bibinfo {author} {\bibfnamefont {K.}~\bibnamefont
  {Liu}}, \bibinfo {author} {\bibfnamefont {J.~Hunter}\ \bibnamefont {Dunn}},
  \bibinfo {author} {\bibfnamefont {K.~M.}\ \bibnamefont {Chesnel}}, \bibinfo
  {author} {\bibfnamefont {J.~B.}\ \bibnamefont {Kortright}}, \bibinfo {author}
  {\bibfnamefont {O.}~\bibnamefont {Hellwig}}, \ and\ \bibinfo {author}
  {\bibfnamefont {E.~E.}\ \bibnamefont {Fullerton.}},\ }\bibfield  {title}
  {\enquote {\bibinfo {title} {Disorder-induced microscopic magnetic memory},}\
  }\href {\doibase 10.1103/PhysRevLett.94.017202} {\bibfield  {journal}
  {\bibinfo  {journal} {Physical Review Letters}\ }\textbf {\bibinfo {volume}
  {94}},\ \bibinfo {pages} {017202} (\bibinfo {year} {2005})}\BibitemShut
  {NoStop}%
\bibitem [{\citenamefont {Malik}\ \emph {et~al.}(1998)\citenamefont {Malik},
  \citenamefont {Sandy}, \citenamefont {Lurio}, \citenamefont {Stephenson},
  \citenamefont {Mochrie}, \citenamefont {McNulty},\ and\ \citenamefont
  {Sutton}}]{MalikPRL1998}%
  \BibitemOpen
  \bibfield  {author} {\bibinfo {author} {\bibfnamefont {A.}~\bibnamefont
  {Malik}}, \bibinfo {author} {\bibfnamefont {A.~R.}\ \bibnamefont {Sandy}},
  \bibinfo {author} {\bibfnamefont {L.~B.}\ \bibnamefont {Lurio}}, \bibinfo
  {author} {\bibfnamefont {G.~B.}\ \bibnamefont {Stephenson}}, \bibinfo
  {author} {\bibfnamefont {S.~G.~J.}\ \bibnamefont {Mochrie}}, \bibinfo
  {author} {\bibfnamefont {I.}~\bibnamefont {McNulty}}, \ and\ \bibinfo
  {author} {\bibfnamefont {M.}~\bibnamefont {Sutton}},\ }\bibfield  {title}
  {\enquote {\bibinfo {title} {Coherent x-ray study of fluctuations during
  domain coarsening},}\ }\href {\doibase 10.1103/PhysRevLett.81.5832}
  {\bibfield  {journal} {\bibinfo  {journal} {Physical Review Letters}\
  }\textbf {\bibinfo {volume} {81}},\ \bibinfo {pages} {5832--5835} (\bibinfo
  {year} {1998})}\BibitemShut {NoStop}%
\bibitem [{\citenamefont {Livet}\ \emph {et~al.}(2001)\citenamefont {Livet},
  \citenamefont {Bley}, \citenamefont {Caudron}, \citenamefont {Geissler},
  \citenamefont {Abernathy}, \citenamefont {Detlefs}, \citenamefont
  {Gr{\"u}bel},\ and\ \citenamefont {Sutton}}]{Livet2001kinetic}%
  \BibitemOpen
  \bibfield  {author} {\bibinfo {author} {\bibfnamefont {F.}~\bibnamefont
  {Livet}}, \bibinfo {author} {\bibfnamefont {F.}~\bibnamefont {Bley}},
  \bibinfo {author} {\bibfnamefont {R.}~\bibnamefont {Caudron}}, \bibinfo
  {author} {\bibfnamefont {E.}~\bibnamefont {Geissler}}, \bibinfo {author}
  {\bibfnamefont {D.}~\bibnamefont {Abernathy}}, \bibinfo {author}
  {\bibfnamefont {C.}~\bibnamefont {Detlefs}}, \bibinfo {author} {\bibfnamefont
  {G.}~\bibnamefont {Gr{\"u}bel}}, \ and\ \bibinfo {author} {\bibfnamefont
  {M.}~\bibnamefont {Sutton}},\ }\bibfield  {title} {\enquote {\bibinfo {title}
  {Kinetic evolution of unmixing in an {AlLi} alloy using x-ray intensity
  fluctuation spectroscopy},}\ }\href {\doibase 10.1103/PhysRevE.63.036108}
  {\bibfield  {journal} {\bibinfo  {journal} {Physical Review E}\ }\textbf
  {\bibinfo {volume} {63}},\ \bibinfo {pages} {036108} (\bibinfo {year}
  {2001})}\BibitemShut {NoStop}%
\bibitem [{\citenamefont {Fluerasu}\ \emph {et~al.}(2005)\citenamefont
  {Fluerasu}, \citenamefont {Sutton},\ and\ \citenamefont
  {Dufresne}}]{fluerasu2005x}%
  \BibitemOpen
  \bibfield  {author} {\bibinfo {author} {\bibfnamefont {Andrei}\ \bibnamefont
  {Fluerasu}}, \bibinfo {author} {\bibfnamefont {Mark}\ \bibnamefont {Sutton}},
  \ and\ \bibinfo {author} {\bibfnamefont {Eric~M.}\ \bibnamefont {Dufresne}},\
  }\bibfield  {title} {\enquote {\bibinfo {title} {X-ray intensity fluctuation
  spectroscopy studies on phase-ordering systems},}\ }\href {\doibase
  10.1103/PhysRevLett.94.055501} {\bibfield  {journal} {\bibinfo  {journal}
  {Physical Review Letters}\ }\textbf {\bibinfo {volume} {94}},\ \bibinfo
  {pages} {055501} (\bibinfo {year} {2005})}\BibitemShut {NoStop}%
\bibitem [{\citenamefont {Kim}\ \emph {et~al.}(2016)\citenamefont {Kim},
  \citenamefont {Kim}, \citenamefont {Jung}, \citenamefont {Kim}, \citenamefont
  {Kim}, \citenamefont {Roth}, \citenamefont {Sprung}, \citenamefont
  {Vartanyants},\ and\ \citenamefont {Ree}}]{kim2016synchrotron}%
  \BibitemOpen
  \bibfield  {author} {\bibinfo {author} {\bibfnamefont {Young~Yong}\
  \bibnamefont {Kim}}, \bibinfo {author} {\bibfnamefont {Kyungtae}\
  \bibnamefont {Kim}}, \bibinfo {author} {\bibfnamefont {Sungmin}\ \bibnamefont
  {Jung}}, \bibinfo {author} {\bibfnamefont {Changsub}\ \bibnamefont {Kim}},
  \bibinfo {author} {\bibfnamefont {Jehan}\ \bibnamefont {Kim}}, \bibinfo
  {author} {\bibfnamefont {Stephan~V}\ \bibnamefont {Roth}}, \bibinfo {author}
  {\bibfnamefont {Michael}\ \bibnamefont {Sprung}}, \bibinfo {author}
  {\bibfnamefont {Ivan~A.}\ \bibnamefont {Vartanyants}}, \ and\ \bibinfo
  {author} {\bibfnamefont {Moonhor}\ \bibnamefont {Ree}},\ }\bibfield  {title}
  {\enquote {\bibinfo {title} {Synchrotron x-ray scattering and photon
  correlation spectroscopy studies on thin film morphology details and
  structural changes of an amorphous-crystalline brush diblock copolymer},}\
  }\href {\doibase 10.1016/j.polymer.2016.08.004} {\bibfield  {journal}
  {\bibinfo  {journal} {Polymer}\ }\textbf {\bibinfo {volume} {105}},\ \bibinfo
  {pages} {472--486} (\bibinfo {year} {2016})}\BibitemShut {NoStop}%
\bibitem [{\citenamefont {Wang}\ \emph {et~al.}(2015)\citenamefont {Wang},
  \citenamefont {Ruta}, \citenamefont {Xiong}, \citenamefont {Zhang},
  \citenamefont {Chushkin}, \citenamefont {Sheng}, \citenamefont {Lou},
  \citenamefont {Cao},\ and\ \citenamefont {Jiang}}]{wang2015free}%
  \BibitemOpen
  \bibfield  {author} {\bibinfo {author} {\bibfnamefont {X.~D.}\ \bibnamefont
  {Wang}}, \bibinfo {author} {\bibfnamefont {B.}~\bibnamefont {Ruta}}, \bibinfo
  {author} {\bibfnamefont {L.~H.}\ \bibnamefont {Xiong}}, \bibinfo {author}
  {\bibfnamefont {D.~W.}\ \bibnamefont {Zhang}}, \bibinfo {author}
  {\bibfnamefont {Y.}~\bibnamefont {Chushkin}}, \bibinfo {author}
  {\bibfnamefont {H.~W.}\ \bibnamefont {Sheng}}, \bibinfo {author}
  {\bibfnamefont {H.~B.}\ \bibnamefont {Lou}}, \bibinfo {author} {\bibfnamefont
  {Q.~P.}\ \bibnamefont {Cao}}, \ and\ \bibinfo {author} {\bibfnamefont
  {J.~Z.}\ \bibnamefont {Jiang}},\ }\bibfield  {title} {\enquote {\bibinfo
  {title} {Free-volume dependent atomic dynamics in beta relaxation pronounced
  {La}-based metallic glasses},}\ }\href {\doibase
  10.1016/j.actamat.2015.08.010} {\bibfield  {journal} {\bibinfo  {journal}
  {Acta Materialia}\ }\textbf {\bibinfo {volume} {99}},\ \bibinfo {pages}
  {290--296} (\bibinfo {year} {2015})}\BibitemShut {NoStop}%
\bibitem [{\citenamefont {Ruta}\ \emph {et~al.}(2017)\citenamefont {Ruta},
  \citenamefont {Zontone}, \citenamefont {Chushkin}, \citenamefont {Baldi},
  \citenamefont {Pintori}, \citenamefont {Monaco}, \citenamefont {Ruffl{\'e}},\
  and\ \citenamefont {Kob}}]{ruta2017hard}%
  \BibitemOpen
  \bibfield  {author} {\bibinfo {author} {\bibfnamefont {Beatrice}\
  \bibnamefont {Ruta}}, \bibinfo {author} {\bibfnamefont {Federico}\
  \bibnamefont {Zontone}}, \bibinfo {author} {\bibfnamefont {Yuriy}\
  \bibnamefont {Chushkin}}, \bibinfo {author} {\bibfnamefont {Giacomo}\
  \bibnamefont {Baldi}}, \bibinfo {author} {\bibfnamefont {Giovanna}\
  \bibnamefont {Pintori}}, \bibinfo {author} {\bibfnamefont {Giulio}\
  \bibnamefont {Monaco}}, \bibinfo {author} {\bibfnamefont {Benoit}\
  \bibnamefont {Ruffl{\'e}}}, \ and\ \bibinfo {author} {\bibfnamefont {Walter}\
  \bibnamefont {Kob}},\ }\bibfield  {title} {\enquote {\bibinfo {title} {Hard
  x-rays as pump and probe of atomic motion in oxide glasses},}\ }\href
  {\doibase 10.1038/s41598-017-04271-x} {\bibfield  {journal} {\bibinfo
  {journal} {Scientific Reports}\ }\textbf {\bibinfo {volume} {7}},\ \bibinfo
  {pages} {3962} (\bibinfo {year} {2017})}\BibitemShut {NoStop}%
\bibitem [{\citenamefont {Bikondoa}(2017)}]{bikondoa2017use}%
  \BibitemOpen
  \bibfield  {author} {\bibinfo {author} {\bibfnamefont {Oier}\ \bibnamefont
  {Bikondoa}},\ }\bibfield  {title} {\enquote {\bibinfo {title} {On the use of
  two-time correlation functions for x-ray photon correlation spectroscopy data
  analysis},}\ }\href {\doibase 10.1107/S1600576717000577} {\bibfield
  {journal} {\bibinfo  {journal} {Journal of Applied Crystallography}\ }\textbf
  {\bibinfo {volume} {50}},\ \bibinfo {pages} {357--368} (\bibinfo {year}
  {2017})}\BibitemShut {NoStop}%
\bibitem [{\citenamefont {Brown}\ \emph {et~al.}(1997)\citenamefont {Brown},
  \citenamefont {Rikvold}, \citenamefont {Sutton},\ and\ \citenamefont
  {Grant}}]{brown1997speckle}%
  \BibitemOpen
  \bibfield  {author} {\bibinfo {author} {\bibfnamefont {Gregory}\ \bibnamefont
  {Brown}}, \bibinfo {author} {\bibfnamefont {Per~Arne}\ \bibnamefont
  {Rikvold}}, \bibinfo {author} {\bibfnamefont {Mark}\ \bibnamefont {Sutton}},
  \ and\ \bibinfo {author} {\bibfnamefont {Martin}\ \bibnamefont {Grant}},\
  }\bibfield  {title} {\enquote {\bibinfo {title} {Speckle from phase-ordering
  systems},}\ }\href {\doibase 10.1103/PhysRevE.56.6601} {\bibfield  {journal}
  {\bibinfo  {journal} {Physical Review E}\ }\textbf {\bibinfo {volume} {56}},\
  \bibinfo {pages} {6601} (\bibinfo {year} {1997})}\BibitemShut {NoStop}%
\bibitem [{\citenamefont {Brown}\ \emph {et~al.}(1999)\citenamefont {Brown},
  \citenamefont {Rikvold}, \citenamefont {Sutton},\ and\ \citenamefont
  {Grant}}]{brown1999evolution}%
  \BibitemOpen
  \bibfield  {author} {\bibinfo {author} {\bibfnamefont {Gregory}\ \bibnamefont
  {Brown}}, \bibinfo {author} {\bibfnamefont {Per~Arne}\ \bibnamefont
  {Rikvold}}, \bibinfo {author} {\bibfnamefont {Mark}\ \bibnamefont {Sutton}},
  \ and\ \bibinfo {author} {\bibfnamefont {Martin}\ \bibnamefont {Grant}},\
  }\bibfield  {title} {\enquote {\bibinfo {title} {Evolution of speckle during
  spinodal decomposition},}\ }\href {\doibase 10.1103/PhysRevE.60.5151}
  {\bibfield  {journal} {\bibinfo  {journal} {Physical Review E}\ }\textbf
  {\bibinfo {volume} {60}},\ \bibinfo {pages} {5151--5162} (\bibinfo {year}
  {1999})}\BibitemShut {NoStop}%
\bibitem [{\citenamefont {Neave}\ \emph {et~al.}(1983)\citenamefont {Neave},
  \citenamefont {Joyce}, \citenamefont {Dobson},\ and\ \citenamefont
  {Norton}}]{neave1983dynamics}%
  \BibitemOpen
  \bibfield  {author} {\bibinfo {author} {\bibfnamefont {J.~H.}\ \bibnamefont
  {Neave}}, \bibinfo {author} {\bibfnamefont {B.~A.}\ \bibnamefont {Joyce}},
  \bibinfo {author} {\bibfnamefont {P.~J.}\ \bibnamefont {Dobson}}, \ and\
  \bibinfo {author} {\bibfnamefont {N.}~\bibnamefont {Norton}},\ }\bibfield
  {title} {\enquote {\bibinfo {title} {Dynamics of film growth of {GaAs} by
  {MBE} from {RHEED} observations},}\ }\href {\doibase 10.1007/BF00617180}
  {\bibfield  {journal} {\bibinfo  {journal} {Applied Physics A}\ }\textbf
  {\bibinfo {volume} {31}},\ \bibinfo {pages} {1--8} (\bibinfo {year}
  {1983})}\BibitemShut {NoStop}%
\bibitem [{\citenamefont {Tsao}(1993)}]{tsao2012materials}%
  \BibitemOpen
  \bibfield  {author} {\bibinfo {author} {\bibfnamefont {Jeffrey~Y.}\
  \bibnamefont {Tsao}},\ }\href
  {https://www.elsevier.com/books/materials-fundamentals-of-molecular-beam-epitaxy/tsao/978-0-08-057135-5}
  {\emph {\bibinfo {title} {Materials fundamentals of molecular beam
  epitaxy}}}\ (\bibinfo  {publisher} {Academic Press},\ \bibinfo {year}
  {1993})\BibitemShut {NoStop}%
\bibitem [{\citenamefont {Kaufmann}\ \emph {et~al.}(2016)\citenamefont
  {Kaufmann}, \citenamefont {Lahourcade}, \citenamefont {Hourahine},
  \citenamefont {Martin},\ and\ \citenamefont
  {Grandjean}}]{kaufmann2016critical}%
  \BibitemOpen
  \bibfield  {author} {\bibinfo {author} {\bibfnamefont {Nils A.~K.}\
  \bibnamefont {Kaufmann}}, \bibinfo {author} {\bibfnamefont {L.}~\bibnamefont
  {Lahourcade}}, \bibinfo {author} {\bibfnamefont {B.}~\bibnamefont
  {Hourahine}}, \bibinfo {author} {\bibfnamefont {D.}~\bibnamefont {Martin}}, \
  and\ \bibinfo {author} {\bibfnamefont {N.}~\bibnamefont {Grandjean}},\
  }\bibfield  {title} {\enquote {\bibinfo {title} {Critical impact of
  {Ehrlich--Schw{\"o}bel} barrier on {GaN} surface morphology during
  homoepitaxial growth},}\ }\href {\doibase 10.1016/j.jcrysgro.2015.06.013}
  {\bibfield  {journal} {\bibinfo  {journal} {Journal of Crystal Growth}\
  }\textbf {\bibinfo {volume} {433}},\ \bibinfo {pages} {36--42} (\bibinfo
  {year} {2016})}\BibitemShut {NoStop}%
\bibitem [{\citenamefont {DenBaars}\ \emph {et~al.}(2013)\citenamefont
  {DenBaars}, \citenamefont {Feezell}, \citenamefont {Kelchner}, \citenamefont
  {Pimputkar}, \citenamefont {Pan}, \citenamefont {Yen}, \citenamefont
  {Tanaka}, \citenamefont {Zhao}, \citenamefont {Pfaff}, \citenamefont
  {Farrell}, \citenamefont {Iza}, \citenamefont {Keller}, \citenamefont
  {Mishra}, \citenamefont {Speck},\ and\ \citenamefont
  {Nakamura}}]{denbaars2013development}%
  \BibitemOpen
  \bibfield  {author} {\bibinfo {author} {\bibfnamefont {Steven~P.}\
  \bibnamefont {DenBaars}}, \bibinfo {author} {\bibfnamefont {Daniel}\
  \bibnamefont {Feezell}}, \bibinfo {author} {\bibfnamefont {Katheryn}\
  \bibnamefont {Kelchner}}, \bibinfo {author} {\bibfnamefont {Siddha}\
  \bibnamefont {Pimputkar}}, \bibinfo {author} {\bibfnamefont {Chi-Chen}\
  \bibnamefont {Pan}}, \bibinfo {author} {\bibfnamefont {Chia-Chen}\
  \bibnamefont {Yen}}, \bibinfo {author} {\bibfnamefont {Shinichi}\
  \bibnamefont {Tanaka}}, \bibinfo {author} {\bibfnamefont {Yuji}\ \bibnamefont
  {Zhao}}, \bibinfo {author} {\bibfnamefont {Nathan}\ \bibnamefont {Pfaff}},
  \bibinfo {author} {\bibfnamefont {Robert}\ \bibnamefont {Farrell}}, \bibinfo
  {author} {\bibfnamefont {Mike}\ \bibnamefont {Iza}}, \bibinfo {author}
  {\bibfnamefont {Stacia}\ \bibnamefont {Keller}}, \bibinfo {author}
  {\bibfnamefont {Umesh}\ \bibnamefont {Mishra}}, \bibinfo {author}
  {\bibfnamefont {James~S.}\ \bibnamefont {Speck}}, \ and\ \bibinfo {author}
  {\bibfnamefont {Shuji}\ \bibnamefont {Nakamura}},\ }\bibfield  {title}
  {\enquote {\bibinfo {title} {Development of gallium-nitride-based
  light-emitting diodes {(LEDs)} and laser diodes for energy-efficient lighting
  and displays},}\ }\href {\doibase 10.1016/j.actamat.2012.10.042} {\bibfield
  {journal} {\bibinfo  {journal} {Acta Materialia}\ }\textbf {\bibinfo {volume}
  {61}},\ \bibinfo {pages} {945--951} (\bibinfo {year} {2013})}\BibitemShut
  {NoStop}%
\bibitem [{\citenamefont {Perret}\ \emph {et~al.}(2017)\citenamefont {Perret},
  \citenamefont {Xu}, \citenamefont {Highland}, \citenamefont {Stephenson},
  \citenamefont {Zapol}, \citenamefont {Fuoss}, \citenamefont {Munkholm},\ and\
  \citenamefont {Thompson}}]{perret2017island}%
  \BibitemOpen
  \bibfield  {author} {\bibinfo {author} {\bibfnamefont {Edith}\ \bibnamefont
  {Perret}}, \bibinfo {author} {\bibfnamefont {Dongwei}\ \bibnamefont {Xu}},
  \bibinfo {author} {\bibfnamefont {M.~J.}\ \bibnamefont {Highland}}, \bibinfo
  {author} {\bibfnamefont {G.~B.}\ \bibnamefont {Stephenson}}, \bibinfo
  {author} {\bibfnamefont {P.}~\bibnamefont {Zapol}}, \bibinfo {author}
  {\bibfnamefont {P.~H.}\ \bibnamefont {Fuoss}}, \bibinfo {author}
  {\bibfnamefont {A.}~\bibnamefont {Munkholm}}, \ and\ \bibinfo {author}
  {\bibfnamefont {Carol}\ \bibnamefont {Thompson}},\ }\bibfield  {title}
  {\enquote {\bibinfo {title} {Island dynamics and anisotropy during vapor
  phase epitaxy of m-plane {GaN}},}\ }\href {\doibase 10.1063/1.4993788}
  {\bibfield  {journal} {\bibinfo  {journal} {Applied Physics Letters}\
  }\textbf {\bibinfo {volume} {111}},\ \bibinfo {pages} {232102} (\bibinfo
  {year} {2017})}\BibitemShut {NoStop}%
\bibitem [{\citenamefont {Pierce}\ \emph {et~al.}(2012)\citenamefont {Pierce},
  \citenamefont {Komanicky}, \citenamefont {Barbour}, \citenamefont {Hennessy},
  \citenamefont {Zhu}, \citenamefont {Sandy},\ and\ \citenamefont
  {You}}]{pierce2012dynamics}%
  \BibitemOpen
  \bibfield  {author} {\bibinfo {author} {\bibfnamefont {Michael~S.}\
  \bibnamefont {Pierce}}, \bibinfo {author} {\bibfnamefont {Vladimir}\
  \bibnamefont {Komanicky}}, \bibinfo {author} {\bibfnamefont {Andi}\
  \bibnamefont {Barbour}}, \bibinfo {author} {\bibfnamefont {Daniel~C.}\
  \bibnamefont {Hennessy}}, \bibinfo {author} {\bibfnamefont {Chenhui}\
  \bibnamefont {Zhu}}, \bibinfo {author} {\bibfnamefont {Alec}\ \bibnamefont
  {Sandy}}, \ and\ \bibinfo {author} {\bibfnamefont {Hoydoo}\ \bibnamefont
  {You}},\ }\bibfield  {title} {\enquote {\bibinfo {title} {Dynamics of the
  {Au} {(001)} surface in electrolytes: in situ coherent x-ray scattering},}\
  }\href {\doibase 10.1103/PhysRevB.86.085410} {\bibfield  {journal} {\bibinfo
  {journal} {Physical Review B}\ }\textbf {\bibinfo {volume} {86}},\ \bibinfo
  {pages} {085410} (\bibinfo {year} {2012})}\BibitemShut {NoStop}%
\bibitem [{\citenamefont {Ju}\ \emph {et~al.}(2017)\citenamefont {Ju},
  \citenamefont {Highland}, \citenamefont {Yanguas-Gil}, \citenamefont
  {Thompson}, \citenamefont {Eastman}, \citenamefont {Zhou}, \citenamefont
  {Brennan}, \citenamefont {Stephenson},\ and\ \citenamefont
  {Fuoss}}]{ju2017instrument}%
  \BibitemOpen
  \bibfield  {author} {\bibinfo {author} {\bibfnamefont {Guangxu}\ \bibnamefont
  {Ju}}, \bibinfo {author} {\bibfnamefont {Matthew~J.}\ \bibnamefont
  {Highland}}, \bibinfo {author} {\bibfnamefont {Angel}\ \bibnamefont
  {Yanguas-Gil}}, \bibinfo {author} {\bibfnamefont {Carol}\ \bibnamefont
  {Thompson}}, \bibinfo {author} {\bibfnamefont {Jeffrey~A.}\ \bibnamefont
  {Eastman}}, \bibinfo {author} {\bibfnamefont {Hua}\ \bibnamefont {Zhou}},
  \bibinfo {author} {\bibfnamefont {Sean~M.}\ \bibnamefont {Brennan}}, \bibinfo
  {author} {\bibfnamefont {G.~Brian}\ \bibnamefont {Stephenson}}, \ and\
  \bibinfo {author} {\bibfnamefont {Paul~H.}\ \bibnamefont {Fuoss}},\
  }\bibfield  {title} {\enquote {\bibinfo {title} {An instrument for in situ
  coherent x-ray studies of metal-organic vapor phase epitaxy of
  {III}-nitrides},}\ }\href {\doibase 10.1063/1.4978656} {\bibfield  {journal}
  {\bibinfo  {journal} {Review of Scientific Instruments}\ }\textbf {\bibinfo
  {volume} {88}},\ \bibinfo {pages} {035113} (\bibinfo {year}
  {2017})}\BibitemShut {NoStop}%
\bibitem [{\citenamefont {Xu}\ \emph {et~al.}(2017)\citenamefont {Xu},
  \citenamefont {Zapol}, \citenamefont {Stephenson},\ and\ \citenamefont
  {Thompson}}]{xu2017kinetic}%
  \BibitemOpen
  \bibfield  {author} {\bibinfo {author} {\bibfnamefont {Dongwei}\ \bibnamefont
  {Xu}}, \bibinfo {author} {\bibfnamefont {Peter}\ \bibnamefont {Zapol}},
  \bibinfo {author} {\bibfnamefont {G.~Brian}\ \bibnamefont {Stephenson}}, \
  and\ \bibinfo {author} {\bibfnamefont {Carol}\ \bibnamefont {Thompson}},\
  }\bibfield  {title} {\enquote {\bibinfo {title} {Kinetic {Monte Carlo}
  simulations of {GaN} homoepitaxy on c-and m-plane surfaces},}\ }\href
  {\doibase 10.1063/1.4979843} {\bibfield  {journal} {\bibinfo  {journal} {The
  Journal of Chemical Physics}\ }\textbf {\bibinfo {volume} {146}},\ \bibinfo
  {pages} {144702} (\bibinfo {year} {2017})}\BibitemShut {NoStop}%
\bibitem [{\citenamefont {Mart{\'i}n-Garc{\'i}a}\ \emph
  {et~al.}(2018)\citenamefont {Mart{\'i}n-Garc{\'i}a}, \citenamefont {Chen},
  \citenamefont {Monta{\~n}a}, \citenamefont {Mascaraque}, \citenamefont
  {Pab{\'o}n}, \citenamefont {Schmid},\ and\ \citenamefont {de~la
  Figuera}}]{2018_Martin-Garcia_SciRep8_5991}%
  \BibitemOpen
  \bibfield  {author} {\bibinfo {author} {\bibfnamefont {Laura}\ \bibnamefont
  {Mart{\'i}n-Garc{\'i}a}}, \bibinfo {author} {\bibfnamefont {Gong}\
  \bibnamefont {Chen}}, \bibinfo {author} {\bibfnamefont {Yaiza}\ \bibnamefont
  {Monta{\~n}a}}, \bibinfo {author} {\bibfnamefont {Arantzazu}\ \bibnamefont
  {Mascaraque}}, \bibinfo {author} {\bibfnamefont {Beatriz~M.}\ \bibnamefont
  {Pab{\'o}n}}, \bibinfo {author} {\bibfnamefont {Andreas~K.}\ \bibnamefont
  {Schmid}}, \ and\ \bibinfo {author} {\bibfnamefont {Juan}\ \bibnamefont
  {de~la Figuera}},\ }\bibfield  {title} {\enquote {\bibinfo {title} {Memory
  effect and magnetocrystalline anisotropy impact on the surface magnetic
  domains of magnetite{(001)}},}\ }\href {\doibase 10.1038/s41598-018-24160-1}
  {\bibfield  {journal} {\bibinfo  {journal} {Scientific Reports}\ }\textbf
  {\bibinfo {volume} {8}},\ \bibinfo {pages} {5991} (\bibinfo {year}
  {2018})}\BibitemShut {NoStop}%
\bibitem [{\citenamefont {Sinha}\ \emph {et~al.}(2014)\citenamefont {Sinha},
  \citenamefont {Jiang},\ and\ \citenamefont {Lurio}}]{sinha2014x}%
  \BibitemOpen
  \bibfield  {author} {\bibinfo {author} {\bibfnamefont {Sunil~K.}\
  \bibnamefont {Sinha}}, \bibinfo {author} {\bibfnamefont {Zhang}\ \bibnamefont
  {Jiang}}, \ and\ \bibinfo {author} {\bibfnamefont {Laurence~B.}\ \bibnamefont
  {Lurio}},\ }\bibfield  {title} {\enquote {\bibinfo {title} {X-ray photon
  correlation spectroscopy studies of surfaces and thin films},}\ }\href
  {\doibase 10.1002/adma.201401094} {\bibfield  {journal} {\bibinfo  {journal}
  {Advanced Materials}\ }\textbf {\bibinfo {volume} {26}},\ \bibinfo {pages}
  {7764--7785} (\bibinfo {year} {2014})}\BibitemShut {NoStop}%
\bibitem [{\citenamefont {Thompson}\ \emph {et~al.}(2008)\citenamefont
  {Thompson}, \citenamefont {Fong}, \citenamefont {Wang}, \citenamefont
  {Jiang}, \citenamefont {Streiffer}, \citenamefont {Latifi}, \citenamefont
  {Eastman}, \citenamefont {Fuoss}, ,\ and\ \citenamefont
  {Stephenson}}]{2008_Thompson_APL93_182901}%
  \BibitemOpen
  \bibfield  {author} {\bibinfo {author} {\bibfnamefont {Carol}\ \bibnamefont
  {Thompson}}, \bibinfo {author} {\bibfnamefont {D.~D.}\ \bibnamefont {Fong}},
  \bibinfo {author} {\bibfnamefont {R.~V.}\ \bibnamefont {Wang}}, \bibinfo
  {author} {\bibfnamefont {F.}~\bibnamefont {Jiang}}, \bibinfo {author}
  {\bibfnamefont {S.~K.}\ \bibnamefont {Streiffer}}, \bibinfo {author}
  {\bibfnamefont {K.}~\bibnamefont {Latifi}}, \bibinfo {author} {\bibfnamefont
  {J.~A.}\ \bibnamefont {Eastman}}, \bibinfo {author} {\bibfnamefont {P.~H.}\
  \bibnamefont {Fuoss}}, , \ and\ \bibinfo {author} {\bibfnamefont {G.~B.}\
  \bibnamefont {Stephenson}},\ }\bibfield  {title} {\enquote {\bibinfo {title}
  {Imaging and alignment of nanoscale {180$^o$} stripe domains in ferroelectric
  thin films},}\ }\href {\doibase 10.1063/1.3013512} {\bibfield  {journal}
  {\bibinfo  {journal} {Applied Physics Letters}\ }\textbf {\bibinfo {volume}
  {93}},\ \bibinfo {pages} {182901} (\bibinfo {year} {2008})}\BibitemShut
  {NoStop}%
\bibitem [{\citenamefont {Mattoni}\ \emph {et~al.}(2016)\citenamefont
  {Mattoni}, \citenamefont {Zubko}, \citenamefont {Maccherozzi}, \citenamefont
  {van~der Torren}, \citenamefont {Boltje}, \citenamefont {Hadjimichael},
  \citenamefont {Manca}, \citenamefont {Catalano}, \citenamefont {Gibert},
  \citenamefont {Liu}, \citenamefont {Aarts}, \citenamefont {Triscone},
  \citenamefont {Dhesi},\ and\ \citenamefont
  {Caviglia}}]{2016_Mattoni_NatCom7_13141}%
  \BibitemOpen
  \bibfield  {author} {\bibinfo {author} {\bibfnamefont {G.}~\bibnamefont
  {Mattoni}}, \bibinfo {author} {\bibfnamefont {P.}~\bibnamefont {Zubko}},
  \bibinfo {author} {\bibfnamefont {F.}~\bibnamefont {Maccherozzi}}, \bibinfo
  {author} {\bibfnamefont {A.~J.~H.}\ \bibnamefont {van~der Torren}}, \bibinfo
  {author} {\bibfnamefont {D.~B.}\ \bibnamefont {Boltje}}, \bibinfo {author}
  {\bibfnamefont {M.}~\bibnamefont {Hadjimichael}}, \bibinfo {author}
  {\bibfnamefont {N.}~\bibnamefont {Manca}}, \bibinfo {author} {\bibfnamefont
  {S.}~\bibnamefont {Catalano}}, \bibinfo {author} {\bibfnamefont
  {M.}~\bibnamefont {Gibert}}, \bibinfo {author} {\bibfnamefont
  {Y.}~\bibnamefont {Liu}}, \bibinfo {author} {\bibfnamefont {J.}~\bibnamefont
  {Aarts}}, \bibinfo {author} {\bibfnamefont {J.-M.}\ \bibnamefont {Triscone}},
  \bibinfo {author} {\bibfnamefont {S.~S.}\ \bibnamefont {Dhesi}}, \ and\
  \bibinfo {author} {\bibfnamefont {A.~D.}\ \bibnamefont {Caviglia}},\
  }\bibfield  {title} {\enquote {\bibinfo {title} {Striped nanoscale phase
  separation at the metal-insulator transition of heteroepitaxial
  nickelates},}\ }\href {\doibase 10.1038/ncomms13141} {\bibfield  {journal}
  {\bibinfo  {journal} {Nature Communications}\ }\textbf {\bibinfo {volume}
  {7}},\ \bibinfo {pages} {13141} (\bibinfo {year} {2016})}\BibitemShut
  {NoStop}%
\bibitem [{\citenamefont {Ravy}(2013)}]{ravy2013homometry}%
  \BibitemOpen
  \bibfield  {author} {\bibinfo {author} {\bibfnamefont {Sylvain}\ \bibnamefont
  {Ravy}},\ }\bibfield  {title} {\enquote {\bibinfo {title} {Homometry in the
  light of coherent beams},}\ }\href {\doibase 10.1107/S0108767313022733}
  {\bibfield  {journal} {\bibinfo  {journal} {Acta Crystallographica Section A:
  Foundations of Crystallography}\ }\textbf {\bibinfo {volume} {69}},\ \bibinfo
  {pages} {543--548} (\bibinfo {year} {2013})}\BibitemShut {NoStop}%
\bibitem [{\citenamefont {Hettel}(2014)}]{hettel2014dlsr}%
  \BibitemOpen
  \bibfield  {author} {\bibinfo {author} {\bibfnamefont {Robert}\ \bibnamefont
  {Hettel}},\ }\bibfield  {title} {\enquote {\bibinfo {title} {{DLSR} design
  and plans: an international overview},}\ }\href {\doibase
  10.1107/S1600577514011515} {\bibfield  {journal} {\bibinfo  {journal}
  {Journal of Synchrotron Radiation}\ }\textbf {\bibinfo {volume} {21}},\
  \bibinfo {pages} {843--855} (\bibinfo {year} {2014})}\BibitemShut {NoStop}%
\bibitem [{\citenamefont {Ju}\ \emph {et~al.}(2018)\citenamefont {Ju},
  \citenamefont {Highland}, \citenamefont {Thompson}, \citenamefont {Eastman},
  \citenamefont {Fuoss}, \citenamefont {Zhou}, \citenamefont {Dejus},\ and\
  \citenamefont {Stephenson}}]{ju2018characterization}%
  \BibitemOpen
  \bibfield  {author} {\bibinfo {author} {\bibfnamefont {Guangxu}\ \bibnamefont
  {Ju}}, \bibinfo {author} {\bibfnamefont {Matthew~J.}\ \bibnamefont
  {Highland}}, \bibinfo {author} {\bibfnamefont {Carol}\ \bibnamefont
  {Thompson}}, \bibinfo {author} {\bibfnamefont {Jeffrey~A.}\ \bibnamefont
  {Eastman}}, \bibinfo {author} {\bibfnamefont {Paul~H.}\ \bibnamefont
  {Fuoss}}, \bibinfo {author} {\bibfnamefont {Hua}\ \bibnamefont {Zhou}},
  \bibinfo {author} {\bibfnamefont {Roger}\ \bibnamefont {Dejus}}, \ and\
  \bibinfo {author} {\bibfnamefont {G.~Brian}\ \bibnamefont {Stephenson}},\
  }\bibfield  {title} {\enquote {\bibinfo {title} {Characterization of the
  x-ray coherence properties of an undulator beamline at the {Advanced Photon
  Source}},}\ }\href {https://arxiv.org/abs/1802.05675} {\bibfield  {journal}
  {\bibinfo  {journal} {Preprint at https://arxiv.org/abs/1802.05675}\ }
  (\bibinfo {year} {2018})}\BibitemShut {NoStop}%
\bibitem [{\citenamefont {Plimpton}\ \emph {et~al.}(2009)\citenamefont
  {Plimpton}, \citenamefont {Battaile}, \citenamefont {Chandross},
  \citenamefont {Holm}, \citenamefont {Thompson}, \citenamefont {Tikare},
  \citenamefont {Wagner}, \citenamefont {Zhou}, \citenamefont
  {Garcia~Cardona},\ and\ \citenamefont
  {Slepoy}}]{2009_Plimpton_SAND2009-6226}%
  \BibitemOpen
  \bibfield  {author} {\bibinfo {author} {\bibfnamefont {Steve}\ \bibnamefont
  {Plimpton}}, \bibinfo {author} {\bibfnamefont {Corbet}\ \bibnamefont
  {Battaile}}, \bibinfo {author} {\bibfnamefont {Mike}\ \bibnamefont
  {Chandross}}, \bibinfo {author} {\bibfnamefont {Liz}\ \bibnamefont {Holm}},
  \bibinfo {author} {\bibfnamefont {Aidan}\ \bibnamefont {Thompson}}, \bibinfo
  {author} {\bibfnamefont {Veena}\ \bibnamefont {Tikare}}, \bibinfo {author}
  {\bibfnamefont {Ed}~\bibnamefont {Wagner}, \bibfnamefont {Greg~andWebb}},
  \bibinfo {author} {\bibfnamefont {Xiaowang}\ \bibnamefont {Zhou}}, \bibinfo
  {author} {\bibfnamefont {Cristina}\ \bibnamefont {Garcia~Cardona}}, \ and\
  \bibinfo {author} {\bibfnamefont {Alex}\ \bibnamefont {Slepoy}},\ }\href
  {http://spparks.sandia.gov/papers.html} {\emph {\bibinfo {title} {Crossing
  the Mesoscale No-Man’s Land via Parallel Kinetic {Monte} {Carlo}}}},\
  \bibinfo {type} {Tech. Rep.}\ \bibinfo {number} {SAND2009–6226}\ (\bibinfo
  {institution} {Sandia National Laboratories},\ \bibinfo {year}
  {2009})\BibitemShut {NoStop}%
\end{thebibliography}%

\section*{Acknowledgments}
We thank Mark Sutton for suggesting the smoothing method used in the speckle analysis, and Dmitry Byelov of ASI and Russell Woods of the APS Detector Pool for expert assistance with the area detector. 
Support provided by the Department of Energy, Office of Science, Basic Energy Sciences, Materials Sciences and Engineering (XPCS measurements and analysis) and Scientific User Facilities (KMC model development). 
Measurements were carried out at the Advanced Photon Source, a DOE Office of Science user facility operated by Argonne National Laboratory. 
Computing resources were provided on Blues and Fusion, high-performance computing clusters operated by the Laboratory Computing Resource Center at Argonne National Laboratory. 
  
\section*{Author Contributions}

G.J., M.J.H., J.A.E, P.H.F., G.B.S., C.T., H.Z., and H.K. developed the pink beam XPCS setup and carried out the measurements.
D.X., P.Z., G.B.S., and C.T. developed and carried out the simulations.
G.J. and G.B.S. analyzed the results.
All coauthors contributed to drafting and editing the manuscript.
  
\section*{Additional Information}
Supplementary Information accompanies this paper. Correspondence and requests for materials or data should be addressed to G.B.S.

\end{document}